\documentclass[letter,11pt]{article}

\pdfoutput=1 

\usepackage{jheppub}
\usepackage{amssymb,amsmath}
\usepackage{graphicx}
\usepackage{epstopdf,epsfig}
\usepackage[scriptsize]{subfigure}
\usepackage[export]{adjustbox}
\usepackage[T1]{fontenc}
\usepackage{xcolor}
\usepackage{marginnote}
\usepackage{mathrsfs}
\usepackage{mathalfa}
\usepackage{upgreek,textgreek}
\usepackage{comment}
\usepackage{floatrow}

\usepackage[color]{changebar}
\cbcolor{red}

\def\imagetop#1{\vtop{\null\scriptsize\hbox{#1}}}

\usepackage{amssymb,amsmath}
\usepackage{epsfig}

\DeclareMathOperator\arctanh{arctanh}
\DeclareMathOperator\arccoth{arccoth}

\newcommand{\be}{\begin{equation}}
\newcommand{\ee}{\end{equation}}
\newcommand{\bea}{\begin{eqnarray}}
\newcommand{\eea}{\end{eqnarray}}
\newcommand{\ba}{\begin{equation}\begin{aligned}}
\newcommand{\ea}{\end{aligned}\end{equation}}

\newcommand{\beg}{\begin{gather*}}
\newcommand{\eng}{\end{gather*}}
\newcommand{\hh}{,\hspace{1cm}}
\newcommand{\hhh}{,\hspace{0.5cm}}

\newcommand{\eq}[1]{(\ref{#1})}

\newcommand{\n}[1]{\label{#1}}

\newcommand{\ins}[1]{{\mbox{\tiny #1}}}

\newcommand{\inds}[1]{{\scriptscriptstyle #1}}
\newcommand{\const}{\mbox{const}}

\newcommand{\ts}[1]{{\boldsymbol{#1}}}

\def\XXint#1#2#3{{\setbox0=\hbox{$#1{#2#3}{\int}$ }
\vcenter{\hbox{$#2#3$ }}\kern-.6\wd0}}

\newcommand{\T}{\tilde{\tau}}
\newcommand{\V}{\tilde{v}}

\title{Two-dimensional black holes in the limiting curvature theory of gravity}

\author{Valeri P. Frolov  and}

\author[1]{Andrei Zelnikov\note{Corresponding author.}}

\affiliation{Theoretical Physics Institute, University of Alberta, Edmonton, Alberta, Canada T6G 2E1}
\emailAdd{vfrolov@ualberta.ca}
\emailAdd{zelnikov@ualberta.ca}

\abstract{
In this paper we discuss modified gravity models in which growth of the curvature is dynamically restricted. To illustrate interesting features of such models we consider a modification of two-dimensional dilaton gravity theory which satisfies the limiting curvature condition. We show that such a model describes two-dimensional black holes which contain the de Sitter-like core instead of the singularity of the original non-modified theory. In the second part of the paper we study Vaidya type solutions of the model of the limiting curvature theory of gravity and used them to analyse properties of black holes which are created by the collapse of null fluid. We also apply these solutions to study interesting features of a black hole evaporation.

}

\begin{document}
\maketitle
\flushbottom

\vspace{-0.5cm}

\section{Introduction}\label{SecI}

Vacuum and electrically charged solutions of the Einstein equations describing stationary black holes have curvature singularities in their interior. In particular  quadratic in curvature invariants infinitely grow in the region close to the singularity. Physically this means that tidal forces become so strong that all physical devices, such as clocks and rulers, which are used for a classical description of the spacetime would be broken in the domain sufficiently close to the curvature singularity. As a result one cannot believe anymore predictions of  the classical Einstein equations.
Famous singularity theorem proved by Penrose \cite{Penrose:1964wq} and later generalized by Penrose and Hawking \cite{Hawking:1969sw,Hawking:1973uf}
implies that this is a generic property of the Einstein gravity, at least in case when the matter satisfies physically reasonable energy conditions.

Singularities inside black holes are somehow similar to the singularities in cosmology. However, there exists an important difference. To explain it let us remind that the Riemann curvature tensor $R_{\alpha\beta\gamma\delta}$ can be decomposed into three parts which include the Weyl tensor $C_{\alpha\beta\gamma\delta}$, traceless Ricci tensor $S_{\alpha\beta}=R_{\alpha\beta}-{1\over 2}g_{\alpha\beta}R$, and Ricci scalar R, respectively. The Kretschmann invariant $\ts{K}=R_{\alpha\beta\gamma\delta}R^{\alpha\beta\gamma\delta}$, which characterizes the curvature strength, can be written in the form
\be
\ts{K}=C^2+2S^2+{1\over 6}R^2\, .
\ee
In the standard homogeneous isotopic cosmological models the  Weyl tensor vanishes.
Then only the Ricci tensor enters the Kretschmann  invariant and, hence, it is enough to properly modify the equation of state of matter to prevent a formation of a cosmological singularity. Such a modification is not sufficient to prevent formation of a black hole singularity. For a vacuum spherically symmetric black hole $S=R=0$ and the growth of the curvature is a consequence of the growth of the Weyl tensor describing the spacetime anisotropy.

It is generally believed that this  ultraviolet incompleteness of the classical  general relativity can be cured by modifying the gravity equations in the high-curvature regime.
The question is what kind of modified gravity (maybe with some extra fields included) does possess the  property that curvature singularities are absent. In such a theory the curvature would not infinitely grow and remains finite. In 1982 Markov formulated  a limiting curvature principle \cite{Markov:1982,Markov:1984ii}. Later similar ideas were discussed in  \cite{Polchinski:1989ae,Morgan:1990yy}. Following his arguments we can propose the following limiting curvature condition:  there should exist a fundamental length scale $\ell$ so that
\be\n{aa1}
{\cal R}\le {\cal B}\,\ell^{-2}\, .
\ee
Here ${\cal R}$ is a scalar invariant describing spacetime curvature. For example, one can choose ${\cal R}=|\ts{K}|^\frac{1}{2}$, or impose a set of such restrictions on the quantities $|C^2|^\frac{1}{2}$, $|S^2|^\frac{1}{2}$ and $|R|$ \ \footnote{It is interesting that in a spherically symmetric spacetime quadratic curvature invariants $C^2$ and $S^2$ are nonnegative, so that in order to impose restrictions on them it is sufficient to impose restriction on the Kretschmann invariant only.}. We also require that dimensionless constant $ {\cal B}$ is universal, that is it depends only on the theory, but it is independent of a particular choice of a solution. In application to black holes this means that
the constant $ {\cal B}$ depends on the choice of the curvature invariant but it does not depend on the mass of the black hole. If condition (\ref{aa1}) is satisfied the curvature singularity in the black hole interior is not formed. We call such objects nonsingular black holes. One can also expect that if the limiting curvature condition is valid the problem of the mass inflation \cite{Poisson:1989zz} can be effectively solved.

Nonsingular black holes have been widely discussed in the literature (see e.g. \cite{Frolov:2016pav,Carballo_Rubio_2020,Rubio:2021obb} and references therein). Such models allow one to solve the information loss problem \cite{Frolov:2014jva} and give hints concerning the final stage of the black hole evaporation. In such models there exists an intriguing possibility of a new universe formation inside the black hole \cite{Poisson_1988,Frolov:1988vj,Frolov:1989pf,Poisson:1990eh,Easson_2001,Dymnikova_2003,Easson:2017pfe,Bardeen:2018frm,Brandenberger:2021ken,Brandenberger:2021jqs}.

Properties of nonsingular black holes are often discussed by postulating the form of the metric \cite{Hayward_2006,Ghosh:2014pba,Frolov:2016pav}. However a real challenge is to find such a consistent modified gravity theory which guarantees the validity of the limiting curvature condition. There were several attempts in this direction \cite{Trodden_1993,Mikovic:1996bh,Easson:2002tg,Kunstatter:2015vxa,Chamseddine:2016ktu,Chamseddine:2019pux}. In this paper we propose a new approach to this problem.

The main idea of this approach is the following. If one has a dynamical system with constraints $Q_i=0$   one can modify the action by adding to its Lagrangian these constraints with Lagrangian multipliers $\sum_i\chi_i Q_i$. A similar method exists for systems when imposed constraints are given by inequalities (see e.g. \cite{1886529043}).
Suppose the imposed inequalities are written in the form $Q_i\le 0$.
Then one can add to the Lagrangian a term $\sum_i\chi_i (Q_i+{\zeta_i}^2)$. By varying the action with respect to the auxiliary variables $(\chi_i,{\zeta_i})$ one gets equations
\be\label{Qi}
Q_i+{\zeta_i}^2=0\hh {\zeta_i} \chi_i=0.
\ee
When $Q_i< 0$ the first equation determines a non-vanishing value of ${\zeta_i}$, while the second equation implies that $\chi_i=0$. However, when $Q_i$ reaches its maximal value, the function ${\zeta_i}$ vanishes while the Lagrange multiplier $\chi_i$ can be nonzero. In this regime some of the auxiliary fields $\chi_i$ become dynamical and enter the equations for the dynamical variables of the original Lagrangian. Mathematical foundations and proof of the related results can be found in the mathematical literature (see e.g. \cite{1886529043}). Certainly, in application to concrete physical problems this method should be properly adjusted.

The aim of this paper is to illustrate how this approach works for a simple two-dimensional dilaton gravity model.
The paper is organized as follows. In section \ref{SecII} we present an action for a two-dimensional dilaton gravity in the limiting curvature theory and derive the field equations. Static solutions describing an eternal black hole in the limiting gravity model are obtained and analysed in sections \ref{SecIII}-\ref{SecV}. Section \ref{SecVI} describes Vaidya type generalization of the static solution. These solutions are used to discuss interesting aspects of the model describing formation and subsequent evaporation of the black hole. The obtained results are discussed in section \ref{SecVII}. The paper has three appendices where some technical details are collected. Appendix~\ref{AppA} contains a derivation of the matching conditions on a junction surface separating the sub- and supercritical domains. A conformal diagram for the metric in the supercritical domain and its matching with a similar diagram for the metric in the subcritical  domain are discussed in appendix~\ref{AppB}. Some useful properties of quantum averages of a 2D conformal massless field in a static 2D black hole background are presented in appendix~\ref{AppC}.


\section{2D Limiting curvature model}\label{SecII}

\subsection{Action}

We begin with the action of a 2D dilaton gravity model \cite{Witten:1991yr} which is known to have black-hole solutions and is exactly solvable at the classical level
\ba\label{Witten1}
 I_\ins{DG}=\frac{1}{2}
    \int d^2x~ |g|^{1/2}~e^{-2\phi}\Big(R+4(\nabla\phi)^2+4\lambda^2\Big).
\ea
This model naturally appears in the framework of string theory and its properties have been extensively studied \cite{Callan:1985ia,Mandal:1991tz,Elitzur:1991cb,Frolov:1992xx,Dijkgraaf:1991ba,Callan:1992rs}.

It is convenient to redefine the dilaton field $\phi$
\ba\label{psiphi}
\psi=e^{-\phi},
\ea
and to rewrite the traditional form \eq{Witten1} of the two-dimensional dilaton gravity (DG) action as follows
\ba\label{Witten2}
 I_\ins{DG}=\frac{1}{2}
    \int d^2x~ |g|^{1/2}~\Big(\psi^2 R+4(\nabla\psi)^2+4\lambda^2\psi^2\Big).
\ea
Our limiting curvature model is obtained  by adding to $ I_\ins{DG}$ an action
\ba\label{I1}
I_\inds{\chi}=\frac{1}{2}\int_{\cal M}d^2x~ |g|^{1/2}~\tilde{\chi}\big( R-\Lambda+\zeta^2\big),\\
\ea
with two Lagrange multipliers $\tilde{\chi}$ and $\zeta$, such that
variation over these parameters would lead to constraints on the curvature. Here $\Lambda>0$ is a positive constant constraining the  2D curvature $R$. Without the Lagrange multiplier $\zeta$ the action \eq{I1} looks exactly as that of  Jackiw–Teitelboim gravity model\cite{Jackiw:1984je,Teitelboim:1983ux}.

Combining actions \eq{Witten2} and \eq{I1} together and redefining Lagrange multiplier $\tilde{\chi}=\chi+\psi^2$ we arrive at the following limiting curvature gravity model $I_\ins{Lim}=I_\ins{DG}+I_\inds{\chi}$
\ba\n{LIM}
 I_\ins{Lim}=&\frac{1}{2}
    \int d^2x\,|g|^{1/2}\Big[\chi R+4(\nabla\psi)^2+4\lambda^2\psi^2
    +(\psi^2-\chi)(\Lambda-{\zeta}^2)\Big].
\ea
It describes the dynamics of a 2D metric $g_{\alpha\beta}$, the dilaton $\psi$, and two Lagrange multiplier fields $\chi$ and $\zeta$.


\subsection{Field equations}

Variation of the action (\ref{LIM}) over $\zeta$ and $\chi$ gives the following constraint equations
\bea\label{constr1}
&(\chi-\psi^2)\zeta=0 \, ,\\
&R-\Lambda+\zeta^2=0 .\label{constr2}
\eea
Variation of the action (\ref{LIM}) over $\psi$ leads to the dilaton field equation
\ba\label{eqpsi1}
\Box\psi-\Big(\lambda^2+\frac{1}{4}(\Lambda-\zeta^2)\Big)\psi=0 .
\ea
Finally variation over $g_{\alpha\beta}$ gives gravity equations which we write in the form
\bea
&&{\cal G}_{\alpha\beta}={\cal T}_{\alpha\beta}\, ,\label{eqchi1}\\
&& {\cal G}_{\alpha\beta}=\chi_{;\alpha\beta}-g_{\alpha\beta}\Box \chi-\frac{1}{2}g_{\alpha\beta}(\Lambda-\zeta^2)\chi\, ,\nonumber\\
&&{\cal T}_{\alpha\beta}=4\psi_{;\alpha}\psi_{;\beta}
-\frac{1}{2}g_{\alpha\beta}\big[ 4\psi_{;\epsilon}\psi^{;\epsilon}\!+\!(4\lambda^2\!+\!\Lambda\!-\!\zeta^2)\psi^2\big].
\nonumber
\eea
Using the constraint \eq{constr2} the gravity equations \eq{eqchi1} can be rewritten in the form
\bea\label{eqchi2}
&&\chi_{;\alpha\beta}+\frac{1}{2}g_{\alpha\beta}R\chi=Q_{\alpha\beta}\, ,\\
&&Q_{\alpha\beta}=4\psi_{;\alpha}\psi_{;\beta}
\!+\!\frac{1}{2}g_{\alpha\beta}\big[\!-\! 4\psi_{;\epsilon}\psi^{;\epsilon}\!+\!(4\lambda^2+R)\psi^2\big].\label{QQ}
\eea
Note that in derivation of \eq{eqchi2}-\eq{QQ} the constraint \eq{constr1} was not used.
It is easy to check that ${\cal G}_{\alpha\beta}$ obeys the ``conservation" law
\be
{\cal G}_{\alpha\beta}^{\ \ ;\beta}=-{1\over 2}R_{;\alpha}\chi\, .
\ee
A similar relation
\be
{\cal T}_{\alpha\beta}^{\ \ ;\beta}=-{1\over 2}R_{;\alpha}\psi^2\,
\ee
is valid for the tensor ${\cal T}_{\alpha\beta}$ provided the other field equations are satisfied.

\subsection{Subcritical and supercritical domains}

The constraint equations (\ref{constr1}) and (\ref{constr2}) imply that there are two different regimes for the solutions.
If the spacetime curvature $R$ is less than its critical value $\Lambda$ then Eq.(\ref{constr2}) determines the field $\zeta$ and this field has a non-vanishing value. We call the solution in such a regime as subcritical. Then Eq.(\ref{constr1}) shows that $\tilde{\chi}=\chi-\psi^2=0$. This means that in this subcritical regime the value of the field $\chi$ is determined by the dilaton field $\psi$.
Thus, to find a solution in the subcritical domain it is sufficient to solve the dilaton equation (\ref{eqpsi1}) and the gravity equation \eq{eqchi1}.
The constraint equation (\ref{constr2}) determines the value of $\zeta$:
\ba
\zeta^2=\Lambda-R\, ,
\ea
while the dilaton and gravity equations, (\ref{eqpsi1}) and  (\ref{eqchi2}), take the form
\ba\label{eqpsi2}
\Box\psi-\Big(\lambda^2+\frac{1}{4}R\Big)\psi=0,
\ea
\ba\label{eqpsi3}
\psi\psi_{;\alpha\beta}-\psi_{;\alpha}\psi_{;\beta}-g_{\alpha\beta}[\Box\psi -\lambda^2\psi]\psi=0.
\ea
Together they lead to
\ba\label{eqpsi4}
\psi\psi_{;\alpha\beta}-\psi_{;\alpha}\psi_{;\beta}=\frac{1}{4}g_{\alpha\beta}R\,\psi^2.
\ea
When the curvature $R$ reaches its maximal value $\Lambda$ the constraint equation (\ref{constr2}) implies that $\zeta=0$. In this supercritical regime the constraint equation (\ref{constr1}) does not impose any restrictions on the field $\chi$ and this fields becomes dynamical. Hence, in the supercritical domain the metric obeys the equation
\be\n{mm}
R=\Lambda\, .
\ee
The fields $\psi$ and $\chi$ obey the equations (\ref{eqpsi1})-(\ref{eqchi2}).

The matching conditions for the fields on junction surface (line) $\Sigma$ where subcritical and supercritical solutions meet  one the other should be found from the field equations.
These matching conditions are discussed in the appendix \ref{AppA}. Their derivation is similar to the approach developed by  Israel's in the general relativity \cite{Israel:1966rt}.

\section{A static black hole: Subcritical domain}\label{SecIII}

\subsection{Solution}

In the case when
\ba
\chi=\psi^2\hh \zeta^2=\Lambda-R
\ea
the constraints are satisfied.
In the parametrization   \eq{psiphi} the gravity equations \eq{eqpsi2},\eq{eqpsi4} reduce to
\bea\label{phiR1}
&&\phi_{;\alpha\beta}=-\frac{1}{4}g_{\alpha\beta}R  ,   \\
\label{phiR2}
&&\phi_{;\alpha}\phi^{;\alpha}=\lambda^2-\frac{1}{4}R .
\eea
In the subcritical domain these equations coincide with the standard equations of 2D dilaton gravity \cite{Witten:1991yr} and their solutions are well known (see e.g. \cite{Fabbri:2005mw}). We briefly remind them in this section mainly in order to fix notations adopted in this paper.

The property \eq{phiR1} guarantees existence of the Killing vector $\xi^{\alpha}=-\varepsilon^{\alpha\beta}\phi_{;\beta}$ \cite{Frolov:1992xx},
where $\varepsilon^{\alpha\beta}$ is 2D Levi-Civita tensor.
We write it in the form $\xi^{\mu}\partial_{\mu}=\partial_t$, where $t$ is the Killing parameter.
The solution of the gravity equations  is  \cite{Witten:1991yr,Frolov:1992xx}
\be \label{Schw}
ds^2=-f dt^2+f^{-1}dr^2\hh
f=1-\frac{m}{\lambda}e^{-2\lambda r},
\ee
while the dilaton is
\ba
\psi=e^{\lambda r}\hh\phi=-\lambda r.
\ea
This is a metric of 2D black hole in an asymptotically flat spacetime. Its horizon is located at
\ba\n{rH}
r_\ins{H}=\frac{1}{2\lambda}\ln\frac{m}{\lambda}
\ea
and the curvature reads
\ba
R=-\frac{\partial^2 f}{\partial r^2}
=4\lambda m \,e^{-2\lambda r} .
\ea
On the horizon $r=r_\ins{H}$  we get $R=4\lambda^2$.

The Killing vector $\ts{\xi}$ is time-like in the black hole exterior, and it is space-like inside the black hole. We normalize it by the condition
\be \n{KIN}
\ts{\xi}^2|_{r=\infty}=-1\, .
\ee
After this the Killing vector is uniquely fixed. As a result, $t$ is a Killing time coordinate outside the black hole horizon, and it is a spatial coordinate inside it.

\subsection{Metric and dilaton field in the black hole interior}

Inside the black hole where $r<r_\ins{H}$ the metric function $f$ is negative.  In this domain  instead of $r$ it is convenient to introduce a new coordinate
\footnote{The subscript $``-"$ stands for objects in the subcritical domain}
$\tau_-$ such that $d\tau_-=-\frac{dr}{\sqrt{-f}}$. Integrating this relation one gets
\ba
&\tau_-=\frac{1}{\lambda}\arctan\sqrt{\frac{m}{\lambda}e^{-2\lambda r}-1} \, .
\ea
The integration constant is chosen so that $\tau_-|_\inds{r=r_\ins{H}}=0$. The inverse transformation reads
\ba
r=\frac{1}{2\lambda}\ln\Big(\frac{m\cos^2(\lambda\tau_{-})}{\lambda}\Big).
\ea
The black-hole metric takes the well known form \cite{Witten:1991yr}
\bea\label{Witten}
&ds^2=-d\tau_{-}^2+a_-^2(\tau_{-})dt^2\, ,\\
&a_-(\tau_{-})=\tan(\lambda\tau_{-})\, .
\eea

The corresponding two-dimensional curvature is
\ba
R=2{\ddot{a}_-\over a_-}=4\frac{\lambda^2}{\cos^2(\lambda\tau_{-})}.
\ea
Here and later we use a dot to denote a derivative with respect to the proper time $\tau$.
The solution for the dilaton reads
\ba
\psi=\sqrt{\frac{m}{\lambda}}\cos(\lambda\tau_{-}).
\ea
The curvature singularity of the solution \eq{Witten} is located at $\tau_{-}=\frac{\pi}{2\lambda}$. In the model with the limiting curvature the solution cannot be extended to this region. Whenever the curvature $R$ reaches its maximum value $\Lambda$ the solution becomes supercritical and has a different form.

The interior metric (\ref{Witten}) describes a two-dimensional expanding universe. Its ``Hubble constant" $H_-$
is
\be
H_-={\dot{a}_-\over a_-}={2\lambda\over \sin(2\lambda \tau_-)}\, .
\ee
It coincides with the extrinsic curvature of a surface (line) $\tau_-=$const (see appendix \ref{AppA}) . We shall use this expression when we shall discuss the matching conditions on the junction surface separating sub- and supercritical domains.

\subsection{Boundary values at the $R=\Lambda$ surface}

The above described solution is valid in the subcritical domain. It should be glued with a solution in the supercritical domain. To formulate the required matching conditions we shall need the boundary values of the metric and the dilaton field at the junction surface $\Sigma$ where the curvature reaches its maximal value $R=\Lambda$. We use subscript $\Lambda$ for these quantities and denote
\ba\label{beta}
\beta=\frac{4\lambda^2}{\Lambda}.
\ea
It is easy to check that
\ba\label{rL}
r_{-,\inds{\Lambda}}=\frac{1}{2\lambda}\ln\frac{4m\lambda}{\Lambda}
\ea
or, equivalently,
\ba\label{tauL}
\tau_{-,\inds{\Lambda}}=\frac{1}{\lambda}\arccos\sqrt{\beta}\, .
\ea
Using (\ref{rH}) one gets
\be\label{rLrH}
r_{-,\inds{\Lambda}}-r_\ins{H}={1\over 2\lambda}\ln\beta\, .
\ee
In what follows we assume that the de Sitter core in located inside the black hole, so that $\beta<1$.

The value of $H_-$ at the junction line is
\be \n{Hm}
H_{-,\inds{\Lambda}}={\lambda \over \sqrt{\beta (1-\beta)}}\, .
\ee
One also has the following expression for the scale factor $a_{-,\inds{\Lambda}}$
\be\n{aaa}
a_{-,\inds{\Lambda}}=\sqrt{ {1-\beta\over \beta}}\, .
\ee
Let us emphasize that the value of the scale parameter $a_{-,\inds{\Lambda}}$ can be changed by a rescaling $t\to \alpha t$ which changes the norm of the Killing vector. The value given in Eq.(\ref{aaa}) is singled out by a chosen normalization condition (\ref{KIN}).

The boundary values of the dilaton field and its derivative on the junction line $\Sigma$ are
\bea
\psi_\inds{\Lambda}&=&\sqrt{ {m\beta\over \lambda}}=2\sqrt{ {m\lambda\over \Lambda}}\, ,\\
\dot{\psi}_\inds{\Lambda}&=&-\sqrt{m\lambda(1-\beta)}\, .
\eea


\subsection{Kruskal coordinates}

To cover a complete spacetime of the 2D black hole we introduce coordinates similar to the standard Kruskal coordinates. For this purpose we start with metric (\ref{Schw}) in the exterior region and write it in the null coordinates
\ba\label{dsuv}
u&={t-r_*}\hhh v={t+r_*}\, .
\ea
Here
\be
r_*=\int {dr\over f}={1\over 2\lambda}\ln \left(\exp(2\lambda r)-{m\over \lambda}\right) .
\ee
In $(u,v)$ coordinates the  metric (\ref{Schw}) takes the form
\ba
&ds^2=-{ du\, dv\over 1+{m\over \lambda} \exp(-\lambda(v-u))} \, .
\ea

As a next step we define new null coordinates $(U,V)$ related with $(u,v)$ as follows
\be
U=-\sqrt{\lambda\over m} \exp(-\lambda u)  \hh
V=\sqrt{\lambda\over m} \exp(\lambda v)\, .
\ee
In these coordinates
\be \n{UV}
ds^2=-{1\over \lambda^2}{dU\, dV\over 1-UV}\, .
\ee
The metric in these coordinates being analytically continued to the domain
\be
-\infty<U<\infty\hh -\infty<V<\infty\hh UV<1\, .
\ee
covers the complete spacetime of the eternal 2D black hole.
At the line $UV=1$ the metric (\ref{UV}) has a curvature singularity.

\begin{figure}[!hbt]
    \centering
    \vspace{10pt}
    \includegraphics[width=0.5 \textwidth]{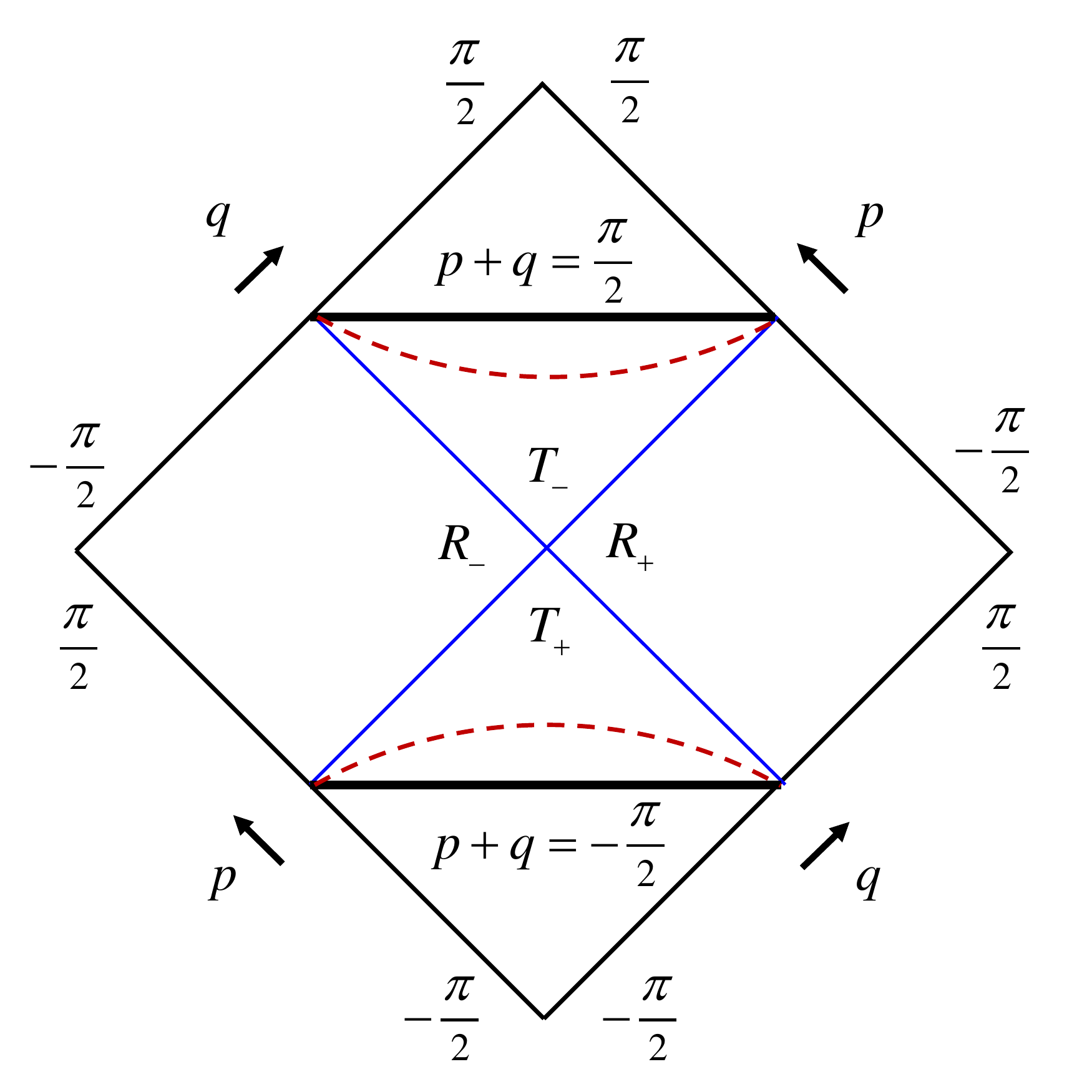}
    \caption{Conformal diagram for the  spacetime of the eternal 2D  black hole. Dashed lines represent junction surfaces where the curvature reaches its maximal value $R=\Lambda$.}
    \label{CPD}
\end{figure}

To construct the Carter-Penrose conformal diagram for the 2D metric  we introduce new null coordinates $(p,q)$
\ba
&U=\tan p \hh -{\pi\over 2}<p<{\pi\over 2} ,\\
&V=\tan q \hh -{\pi\over 2}<q<{\pi\over 2} .
\ea
In these coordinates the metric takes the form
\be\n{pqpq}
ds^2=-\Omega\, dp\, dq\hh
 \Omega={1\over \lambda^2}{1\over \cos p\, \cos q\, \cos (p+q)}\, .
\ee
Lines $p+q=\pm {\pi \over 2}$ correspond to spacelike curvature  singularities. The conformal diagram for the eternal 2D black in these coordinates is depicted in figure~\ref{CPD}.

The causal structure of the spacetime of the eternal 2D black hole is qualitatively similar to the standard black hole. The conformal diagram contains four wedges, which we denote as $R_{\pm}$ and $T_{\pm}$. The metric (\ref{pqpq}) admits the following discrete symmetries:
\be
\Psi_1 : p\to q \hhh q\to p\,; \hspace{2cm}\Psi_2 : p\to- q \hhh q\to - p\, .
\ee
The map $\Psi_1$ interchanges the regions $R_+$ and $R_-$ while preserving $T_{\pm}$ regions.  The map $\Psi_2$ moves $T_-\to T_+$ and $T_+\to T_-$ while preserving the position of $R_{\pm}$. The spacetime of the eternal black hole is empty. If matter is included only part  of it describing the metric outside of the matter is used, while inside the matter the geometry should be found by solving the corresponding equations. In the description of the gravitational collapse and black hole formation it is sufficient to use parts of the domains $R_+$ and $T_-$. For the description of the white hole one uses $T_+$ and $R_+$ domains. The spacetime of the eternal black hole has two spacelike singularities shown by  solid lines in figure~\ref{CPD}. In the limiting curvature theory of gravity there exist two junction surfaces $\Sigma$ shown by dashed lines. However it is sufficient to glue sub- and supercritical solutions at one of this junction lines. In what follows we always assume that the chosen junction line $\Sigma$ is located in the black hole interior, that is in $T_-$ domain. The corresponding junction conditions on the other junction line can be obtained by means of $\Psi_2$ symmetry map.

\section{Solution  in the supercritical domain}\label{SecIV}

\subsection{Metric}

When the curvature reaches the critical value $\Lambda$ the constraint \eq{constr2} requires that $\zeta=0$. This condition also automatically fulfills \eq{constr1}.
Equation (\ref{constr2}) for $\zeta=0$ implies that the curvature in the supercritical domain is constant and equal to $\Lambda$.
This means that the geometry in this domain is isometric to the 2D de Sitter spacetime. We write its metric  in the form
\ba\label{m+}
ds^2=-d\tau_+^2+a(\tau_+)^2 dt^2  \, .
\ea
The de Sitter metric has three Killing vectors. In the domain covered by the coordinates $(\tau_+,t)$ the Killing vector $\xi^{\mu}\partial_{\mu}=\partial_t$ is spacelike and $a^2=\ts{\xi}^2$. This Killing vector is singled out by the property that it is tangent to the junction surface and coincides on it with the Killing vector of the subcritical domain. This condition, which we shall discuss in more detail later, fixes the norm of $\ts{\xi}$ in the supercritical domain and provides one with a unique choice of the Killing coordinate $t$.

Let us emphasize that quite often the metric (\ref{m+}) is used with an assumption that its spatial sections are compact and has topology $S^1$. In such a case the coordinate $t$ is periodic (see e.g  \cite{Spradlin:2001pw} and references therein). We do not assume such a periodicity so that
the coordinate $t$ is chosen to spans the interval $ t \in [-\infty,\infty]$ which is the same as that of the corresponding Killing coordinate in the black hole patch.  Actually in Minkowski signature both choices are legitimate and are used depending on the physical setup of the problem \cite{Ness:2002qr}.

We shall  use the following dimensionless version of $\tau_{+}$
\be
\tilde{\tau}=\sqrt{ {\Lambda\over 2}}\tau_+\, .
\ee
As earlier we denote by a dot a derivative with respect the proper time $\tau_+$, and use a prime to denote a derivative with respect to $\tilde{\tau}$.
The curvature for the metric (\ref{m+}) is
\ba
R=2\frac{\ddot{a}}{a}=\Lambda {a''\over a}\, .
\ea
The solution of the equation $R=\Lambda$ reads
\ba
a=B_1 e^{\T}+B_2 e^{-\T},
\ea
where $B_{1,2}$ are integration constants.
Using freedom in shifting of the coordinate $\tau$ by a constant,
one can write the metric function $a$  in one of the  following forms \footnote{We do not consider the case $a\sim \exp(-\T)$ since as we shall see later it is excluded by the matching conditions for our problem.}
\ba\label{AAA}
a_k=\begin{cases}
&A_1 \cosh(\T),  ~~ \, \mbox{for \ }k=+1 ,\\
&A_{-1}\sinh(\T),  ~ \mbox{for \ }k=-1 ,\\
&A_0\exp(\T), \hspace{0.5cm} \mbox{for \ }k=0.
\end{cases}
\ea
The corresponding quantity $H={\dot{a}\over a}$ for these cases is
\ba
H_k =\sqrt{{\Lambda\over 2}}
\begin{cases}
&\tanh(\T),\ \mbox{for \ }k=+1 ,\\
&\coth(\T),\ \,\mbox{for \ }k=-1 ,\\
&1,  \hspace{1.13cm} \ \mbox{for \ }k=0 .
\end{cases}
\ea

\subsection{Matching conditions}

The matching conditions on the junction surface $\Sigma$ which are required for gluing sub- and supercritical solutions  follow from the field equations. These conditions are derived in appendix~\ref{AppA}. They require that the metric induced on the junction surface $\Sigma$ by its embedding in sub- and supercritical geometry should be the same and the extrinsic curvature is continuous on $\Sigma$. In the case of the eternal black hole there exist two such junction lines. As it was explained earlier because of the symmetry of the eternal black hole metric it is sufficient to glue sub- and supercritical solutions only on one of these surfaces. We chose $\Sigma$ located in the black hole interior, that is in $T_-$ domain.

In the vicinity of the spacelike junction line $\Sigma$, which corresponds to the choice $\epsilon=-1$ (see appendix \ref{AppA}), one can introduce Gaussian normal coordinates in which the metric takes the form
\be \n{m1}
ds^2=-d\tau^2+a^2_{\pm} (\tau)dt_{\pm}^2\, .
\ee
Since $a_{\pm}$ does not depend on $t_{\pm}$ the matching conditions derived in the appendix~\ref{AppA} are greatly simplified.
The subscript $``-"$ stands for coordinates and other objects in the subcritical domain, and the index $"+"$ is used for quantities in the supercritical domain. The equation of $\Sigma$ in these coordinates is $\tau=\tau_0=$const .
The line elements $d\ell^2=a^2_{\pm} dt_{\pm}^2$ on $\Sigma$ induced by its embedding in the 2D metric in both $\pm$ regions should be the same. This gives
\be \n{m2}
a_-(\tau_0) dt_-=a_+(\tau_0) dt_+\, .
\ee
The metrics (\ref{m1}) have a Killing vectors which are spacelike on $\Sigma$. One can identify the parameters $t_{\pm}$ with the Killing parameter $t$, then the condition of continuity of the Killing vector on $\Sigma$ implies
\be
t_{\pm}=t\hh a_-(\tau_0)=a_+(\tau_0)\, .
\ee

The extrinsic curvature of $\Sigma$ is
\be
H_{\pm}={1\over a}{\partial a_{\pm}\over \partial \tau}\, .
\ee
Since it should be continuous on $\Sigma$ one has
\be\n{HHH}
[H]=(H_+ -H_-)_{\tau_0}=0\, .
\ee
The curvature for the metric (\ref{m1}) can be written in the form
\be
R=2\left({\partial H\over \partial \tau}+H^2\right) \, .
\ee
The condition (\ref{HHH}) guarantees that there is no $\delta(\tau)$ contribution to the curvature.

The matching condition for $\zeta$ is $[\zeta]=0$. Equation (\ref{eqpsi1}) implies
\be
[\psi]=[\partial \psi/\partial \tau]=0\, .
\ee

In appendix~\ref{AppA} it is shown that  it is sufficient to require that
\be
[\chi]=0\, .
\ee
then all other matching conditions which involve $\partial_{\tau}\chi$ and $\partial^2_{\tau}\chi$ are satisfied.

\subsection{Gluing sub- and supercritical metrics}

The matching conditions for geometries requires continuity of $H=\dot{a}/a$ on the junction surface $\Sigma$. Thus one has
\ba
H_{-,\inds{\Lambda}}\Big|_{\tau_{-}}=H_k\Big|_{\tau_{+}=\tau_{0}}\, ,
\ea
where $H_{-,\inds{\Lambda}}$ is given by Eq.(\ref{Hm}).
Then the matching  condition takes the form
\ba\label{cases}
\frac{1}{\sqrt{2(1-\beta)}}
=
\begin{cases}
&\tanh(\T_0),\ \mbox{for \ }k=+1 ,\\
&\coth(\T_0),\hspace{0.2cm} \mbox{for \ }k=-1 ,\\
&1, \hspace{1.4cm} \mbox{for \ }k=0 \, ,
\end{cases}
\ea
where
$
\tilde{\tau}_0=\sqrt{ {\Lambda\over 2}}\tau_0\, .
$

The coordinates $\tau_-$ and $\tau_+$ are discontinuous at the junction point. If we want to have a continuous coordinate $\tau$, then we have to shift, e.g., $\tau_+\to \bar{\tau}_+ -\Delta\tau$ by the value of this jump.

To fix a solution in the supercritical domain which satisfies proper matching condition it is sufficient to determine the following parameters: $\tilde{\tau}_0$, $\Delta\tilde{\tau}$ and $A_k$. The form of these expressions depends on $k$  and is slightly different for its different values. Let us discuss these cases.
\begin{enumerate}
\item If $k=+1$, then $0\le\beta<\frac{1}{2}$  and the initial value of $\tau_+=\tau_0$ and the shift $\Delta\tau$ are defined by the  equations
\ba\label{tau01}
&\tanh(\T_0)=\frac{1}{\sqrt{2(1-\beta)}},\\
&\lambda\Delta\tau=\arccos\sqrt{\beta}-\frac{\sqrt{\beta}}{\sqrt{2}}\arccoth(\sqrt{2(1-\beta)}).
\ea
If we require the Killing coordinate $t$ to be continuous on the surface of the transition, then the scale factor $A_1$ in \eq{AAA} is
fixed by the condition that $a_{-,\inds{\Lambda}}=a_+(\tau_0)$.
It leads to
\bea
A_1=\sqrt{\frac{1}{2\beta}-1} \hh a_+(\tau_0)=\sqrt{\frac{1-\beta}{\beta}}.
\eea

\item  If $k=-1$, then $\frac{1}{2}<\beta\le1$ and the equation for $\tau_0$ is
\ba\label{tau02}
\coth(\T_0)=\frac{1}{\sqrt{2(1-\beta)}}.
\ea
The jump $\Delta\tau=\tau_\Lambda-\tau_0$ in the $\tau$ coordinate is
\bea
\lambda\Delta\tau=\arccos\sqrt{\beta}-\frac{\sqrt{\beta}}{\sqrt{2}}\arctanh(\sqrt{2(1-\beta)}) .
\eea
Similarly to the previous case the requirement of continuity of the Killing coordinate $t$ leads to the condition
\bea
A_{-1}=\sqrt{1-\frac{1}{2\beta}}\hh a_+(\tau_0)=\sqrt{\frac{1-\beta}{\beta}}.
\eea
\item  If  $k=0$,
the parameter $\tilde{\tau}_0$ does not enter in the matching condition for $H$.  At the same time the requirement of continuity of $a$ implies that
\be
A_0\exp(\tilde{\tau}_0)=\sqrt{1-\beta\over \beta}\, .
\ee
This relation does not define $A_0$ and $\tilde{\tau}_0$ separately. However, in what follows this is not necessary.
\end{enumerate}

\begin{figure}[!hbt]
    \centering
    \vspace{10pt}
    \includegraphics[width=0.55 \textwidth]{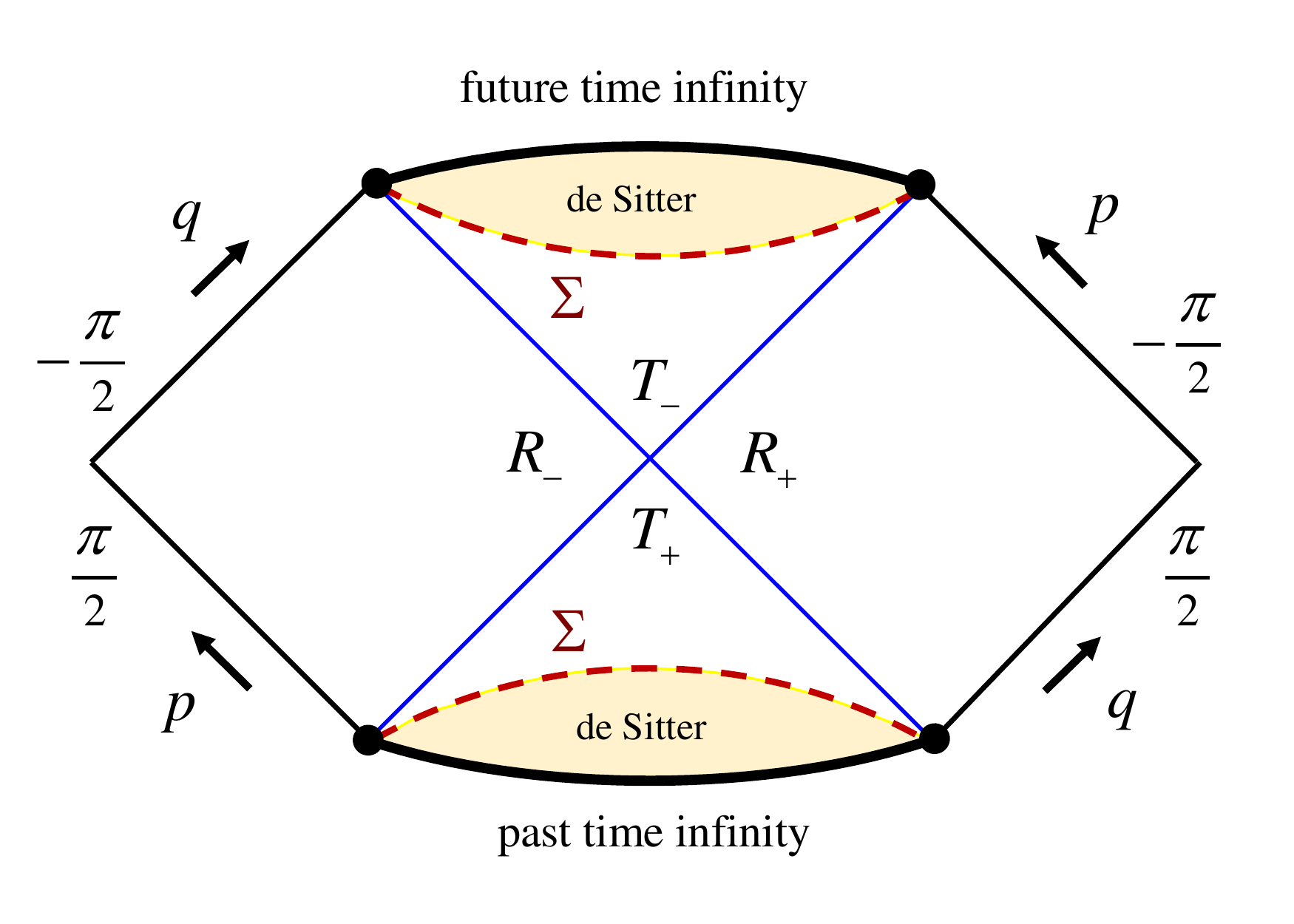}
    \caption{Conformal diagram for the  spacetime of the solution of limiting curvature gravity theory which consists of so to say an `eternal' 2D black hole glued to the de Sitter spacetimes. Dashed lines represent junction surfaces where the curvature reaches its maximal value $R=\Lambda$.}
    \label{CPD4}
\end{figure}

Using the obtained matching conditions one can ``glue" the metrics of the sub- and supercritical domains. In particular, this allows one to construct a conformal diagram for the global solution. The details of this procedure can be found in appendix~\ref{AppB}. Here we just present a schematic conformal diagram for the case $k=+1$ ($\beta<1/2$). This diagram is shown in figure~\ref{CPD4}.

The eternal black hole in the limiting curvature gravity has two de Sitter cores. The first one located in the white hole region $T_+$ presents the contracting de Sitter universe. The second one located in the black hole interior $T_-$ presents the expanding de Sitter universe.

\section{Dilaton field in supercritical domain}

When $\zeta=0$, the constraint equations guarantee that the curvature is constant
\ba\label{E1}
R=\Lambda .
\ea
The dilaton equation then becomes
\ba\label{E2a}
\Box\psi-\lambda^2(1+\frac{1}{\beta})\psi=0  \, ,
\ea
and the gravity equations \eq{eqchi2} take the form
\ba\label{eqchi3}
\chi_{;\alpha\beta}&+\frac{1}{2}g_{\alpha\beta}\Lambda\chi
=4\psi_{;\alpha}\psi_{;\beta}
+\frac{1}{2}g_{\alpha\beta}\big[-4\psi_{;\epsilon}\psi^{;\epsilon}+(4\lambda^2+\Lambda)\psi^2\big].
\ea
Firstly we solve the equation for the dilaton. The solutions look slightly different depending on the value of the limiting curvature $\Lambda$.

\subsubsection{Case $k=+1$}
In this case $ 0\le\beta<\frac{1}{2}$ and the scale factor is given by
\ba\label{a}
a=A_{1}\cosh \T .
\ea
The dilaton equation reads
\ba\n{PPP}
{\psi}''+\tanh(\T){\psi}'+\frac{1}{2}(1+\beta)\psi=0 .
\ea
Two independent real solutions are expressed in terms of hypergeometric functions
\ba\label{2A}
&\psi_1={}_\inds{2\!}F_\ins{1}\Big(\frac{1}{4}+\frac{i\nu}{2},\frac{1}{4}-\frac{i\nu}{2};\frac{1}{2};-z^2\Big),\\
&\psi_2=z\,
{}_\inds{2\!}F_\ins{1}\Big(\frac{3}{4}+\frac{i\nu}{2},\frac{3}{4}-\frac{i\nu}{2};\frac{3}{2};-z^2\Big),
\ea
where
\ba
z=\sinh \T \hh
\nu=\frac{1}{2}\sqrt{1+2\beta}.
\ea
We provide here two more forms of the linear independent solutions of the equation (\ref{PPP}) which are useful of our further considerations.
\begin{itemize}
\item The real basis
\ba\label{2B}
&\widetilde{\psi}_1=\frac{P_{-\frac{1}{2}}^{i\nu}\big(\frac{z}{\sqrt{z^2+1}}\big)
+P_{-\frac{1}{2}}^{-i\nu}\big(\frac{z}{\sqrt{z^2+1}} \big)}{(z^2+1)^{\frac{1}{4}}},\\
&\widetilde{\psi}_2=\frac{Q_{-\frac{1}{2}}^{i\nu}\big(\frac{z}{\sqrt{z^2+1}} \big)}{(z^2+1)^{\frac{1}{4}}}.
\ea
\item The complex basis
\ba\label{2C}
&\bar{\psi}_1=\frac{1}{z^{\frac{1}{2}+i\nu}}~
{}_\ins{2}F_\ins{1}\Big(\frac{1}{4}+\frac{i\nu}{2},\frac{3}{4}+\frac{i\nu}{2};1+i\nu;-\frac{1}{z^2}\Big),\\
&\bar{\psi}_2=\frac{1}{z^{\frac{1}{2}-i\nu}}~
{}_\ins{2}F_\ins{1}\Big(\frac{1}{4}-\frac{i\nu}{2},\frac{3}{4}-\frac{i\nu}{2};1-i\nu;-\frac{1}{z^2}\Big).
\ea
\end{itemize}
The latter form of solutions is specially useful for study the asymptotic behavior of the dilaton field for large $\tau$. In this case
the last arguments of the  hypergeometric functions tends to $0$ and the asymptotic of the solution takes the form
\ba\label{2C}
&\bar{\psi}_1\to 2^{\frac{1}{2}+i\nu}e^{-\T(\frac{1}{2}+i\nu)},
\\
&\bar{\psi}_2\to 2^{\frac{1}{2}-i\nu} e^{-\T(\frac{1}{2}-i\nu)}.
\ea

\subsubsection{Case $k=-1$ }
In this case when $1/2<\beta<1$ the cosmological factor $a$ has the form
\ba\label{a}
a=A_{-1}\sinh(\T).
\ea
Then the dilaton equation can be written explicitly
\ba
{\psi}''+\coth(\T){\psi}'+\frac{1}{2}(1+\beta)\psi=0 .
\ea
In terms of the variable
\ba
z=\cosh(\T)
\ea
two independent complex solutions can be expressed in terms of Legendre functions
\ba\label{1A}
&\psi_1=P_{-\frac{1}{2}+i\nu}\big(z \big),\\
&\psi_2=Q_{-\frac{1}{2}+i\nu}\big(z \big).
\ea
An equivalent pair of real solutions reads
\ba\label{1B}
&\widetilde{\psi}_1=\frac{P_{-\frac{1}{2}}^{i\nu}\big(\frac{z}{\sqrt{z^2-1}}\big)
+P_{-\frac{1}{2}}^{-i\nu}\big(\frac{z}{\sqrt{z^2-1}} \big)}{(z^2-1)^{\frac{1}{4}}},\\
&\widetilde{\psi}_2=\frac{Q_{-\frac{1}{2}}^{i\nu}\big(\frac{z}{\sqrt{z^2-1}} \big)}{(z^2+1)^{\frac{1}{4}}}.
\ea
\subsubsection{Case $k=0$.  }

We have $\beta=\frac{1}{2}$ and the scale factor
\ba
a=A_0e^{\T}.
\ea
The corresponding dilaton equation
\ba
{\psi}''+{\psi}'+\frac{3}{4}\psi=0 .
\ea
has a simple set of solutions
\ba\label{3B}
&\psi_1=e^{-\frac{\T}{2}}\cos\Big(\frac{\T}{\sqrt{2}}\Big)=z^{-1/2}\cos\Big(\frac{\ln z}{\sqrt{2}}\Big),\\
&\psi_2=e^{-\frac{\T}{2}}\sin\Big(\frac{\T}{\sqrt{2}}\Big)=z^{-1/2}\sin\Big(\frac{\ln z}{\sqrt{2}}\Big),
\ea
where $z=e^{\T}$.

Let us emphasize that asymptotic behavior of the dilaton field for large $\tilde{\tau}$ in all three cases $k=\pm 1,0$ is the same. The solutions oscillate while their amplitudes exponentially decrease.


\section{Solving gravity equations}\label{SecV}

\subsection{Solution}

As we already mentioned in the supercritical domain we have singled out one of the symmetry generating Killing vectors by the property that it is tangent to the junction line $\Sigma$ and coincides on it with the Killing vector of the subcritical domain. Since the metric and the fields on $\Sigma$ respect this symmetry they should obey the similar conditions in the supercritical domain. In particular, one has
\be\label{kill}
\xi^{\alpha}\psi_{;\alpha}=\xi^{\alpha}\chi_{;\alpha}=0\, .
\ee
The gravity equations \eq{eqchi2} in the supercritical domain take the form
\be \n{xx}
\chi_{\alpha\beta}+{1\over 2}g_{\alpha\beta}R\chi=Q_{\alpha\beta}\, ,
\ee
where $R=\Lambda$ and $Q_{\alpha\beta}$ is given by Eq.(\ref{QQ}).
Let us contact equation (\ref{xx}) with $\xi^{\alpha}\xi^{\beta}/\ts{\xi}^2$
\ba
&{\cal Q}=
\frac{\xi^{\alpha}\xi^{\beta}}{\ts{\xi}^2}Q_{\alpha\beta}
=-2\psi_{;\epsilon}\psi^{;\epsilon}+\frac{1}{2}\big(4\lambda^2+\Lambda\big)\psi^2\, .
\ea
The Killing vector has the property
\be
\Box \ln (\ts{\xi}^2)=-R \, .
\ee
Using \eq{kill} one  obtains
\be
\frac{\xi^{\alpha}\xi^{\beta}}{\ts{\xi}^2}\chi_{;\alpha\beta}=\frac{1}{2}\chi_{;\alpha}\nabla^\alpha \ln (\ts{\xi}^2) .
\ee
In the metric \eq{m+} we obtain
\bea
&\ts{\xi}^2=a^2 \hh \ln (\ts{\xi}^2)^{;\alpha}=(-2H,0)\hh H=\dot{a}/a \, ,\\
&{\cal Q}=2\dot{\psi}^2+\frac{1}{2}\big(4\lambda^2+\Lambda\big)\psi^2\, .
\eea
Thus we get following first order equation for $\chi$
\be
-\dot{a}\dot{\chi}+\ddot{a}\chi=a{\cal Q}\, .
\ee
Its solution reads
\be\label{chitau}
\chi(\tau)=\dot{a}\Big(c_0-\int_{\tau_0}^{\tau}d\tau \frac{a}{\dot{a}^2} {\cal Q}\Big)\, .
\ee
Here $\tau_0$ is the value of $\tau$ at the junction surface. The constant of integration $c_0$ is fixed by the junction condition
\ba
\chi(\tau_0)=\psi^2(\tau_0).
\ea
It leads to
\ba
c_0=\frac{\psi^2(\tau_0)}{\dot{a}(\tau_0)}.
\ea
Let us introduce a new variable
\ba
z=\frac{1}{A_k \omega}\dot{a}\hh \omega=\sqrt{\frac{\Lambda}{2}}
\ea
instead of $\tau$, then we have $\dot{z}=\omega a/A_k$. The solution \eq{chitau} becomes
\ba\label{chitau1}
&\chi= z\Big\{\omega c_0-\int_{z_0}^{z} \frac{dz}{ z^2} \big(2a^2(\partial_z\psi)^2+(\beta+1)\psi^2\big)\Big\},
\\
&\omega c_0=\frac{\psi^2}{z_0}\Big|_{z=z_0}.
\ea
It is easy to check that one can write the expression for $\chi$ in the following form which is valid for all the cases $k=\pm1,0$
\be\label{chitau1}
\chi= z\Big\{\frac{\psi^2(z_0)}{z_0}-\int_{z_0}^{z} \frac{dz}{ z^2} \big(2(z^2+k)(\partial_z\psi)^2+(\beta+1)\psi^2\big)\Big\}\, .
\ee
The expression for the dilaton field as a function of $z$ which enters this integral
are given in \eq{2B},\eq{1B},\eq{3B}.


\subsection{Can a solution leave the supercritical regime?}

In order to answer this question one can proceed as follows. We denote
\be
Z={\chi-\psi^2\over a'}\, .
\ee
Changing regime condition reads $Z=0$. The following set of equations
determines the dynamics of $Z$
\ba\n{eqq1}
\psi'&=j,\\
j'&=-{a'\over a}j-\frac{1}{2}(1+ \beta)\psi ,\\
Z'&=-{2a\over a'^2}\big( j^2+\frac{\beta}{2}\psi^2+{a'\over a}\psi j\big)
=-{2a\over a'^2}\Big(\big( j+{a'\over 2 a}\psi\big)^2+\big(\frac{\beta}{2}-\frac{a'^2}{4a^2}\big)\psi^2\Big).
\ea
The first two equations of this system are nothing but dilaton equation written in the first order form, while the last equation which determines the evolution of the parameter $Z$ follows from the gravity equation for $\chi$. The initial conditions for this system are
\be
Z=0\hh \psi=2\sqrt{{m\lambda\over \Lambda}}\hh j=-\sqrt{{2m\lambda\over \Lambda}}\sqrt{1-\beta}\, .
\ee

Note that at the initial point $\T_0$ we have $a'/a|_{\T=\T_0}=1/\sqrt{2(1-\beta)}$ and, as the result, we also have $Z'|_{\T=\T_0}=0$.

This system of equations can be further simplified. Let us introduce new variables
\be
\Psi={1\over 2}\sqrt{ {\Lambda\over m\lambda}}\psi\hh
J={1\over 2}\sqrt{ {\Lambda\over m\lambda}} j\hh
\tilde{Z}={\Lambda\over 4m\lambda} Z\, .
\ee
Then the system of equations (\ref{eqq1}) takes the form
\ba\n{eqq3}
&\Psi'=J\, ,\\
&J'=-{a'\over a}J-{1+\beta\over 2}\Psi\, ,\\
&\tilde{Z}'=-{2a\over a'^2}\big( J^2+{\beta\over 2}\Psi^2+{a'\over a}\Psi J\big)\, ,
\ea
while the initial conditions are
\be\label{Zinit}
\tilde{Z}=0\hh \Psi=1\hh J=-{1\over \sqrt{2}}\sqrt{1-\beta}\, .
\ee
Let us emphasize that for the given function $a$  both initial conditions and the system itself written in these new variables depend only on one parameter $\beta$.

Suppose a solution enters into the supercritical regime. It can leave it only if the function $Z$ at some moment of time $\tilde{\tau}$ passes through zero again. Before we discuss this question in details, let us formulate first our conclusion: In the discussed 2D version of the limiting curvature theory of gravity the return of the solution from the supercritical regime to the subcritical one is impossible. For $k=0$ this result can be proved analytically, while for $k=\pm 1$ to obtain this conclusion we performed numerical calculations.

Let us start with the case $k=0$ ( $\beta=\frac{1}{2}$).  For this case
\be
{a'\over a}=\sqrt{\beta}\, .
\ee
This relation implies that the right-hand side of the equation for $Z'$ \eq{eqq1} is always negative.
This means that at least in this case for the given initial conditions (\ref{Zinit}) the function $Z$ is negative for $\tilde{\tau}>\tilde{\tau}_0$ and it cannot pass zero again.

\begin{figure}[!htb]
\centering \setlength{\labelsep}{0.0mm}
\begin{tabular}{l l l l}
\hspace{-0.05in}\imagetop{(a)} & \hspace{-0.1in}\imagetop{\includegraphics[width=0.475\textwidth]{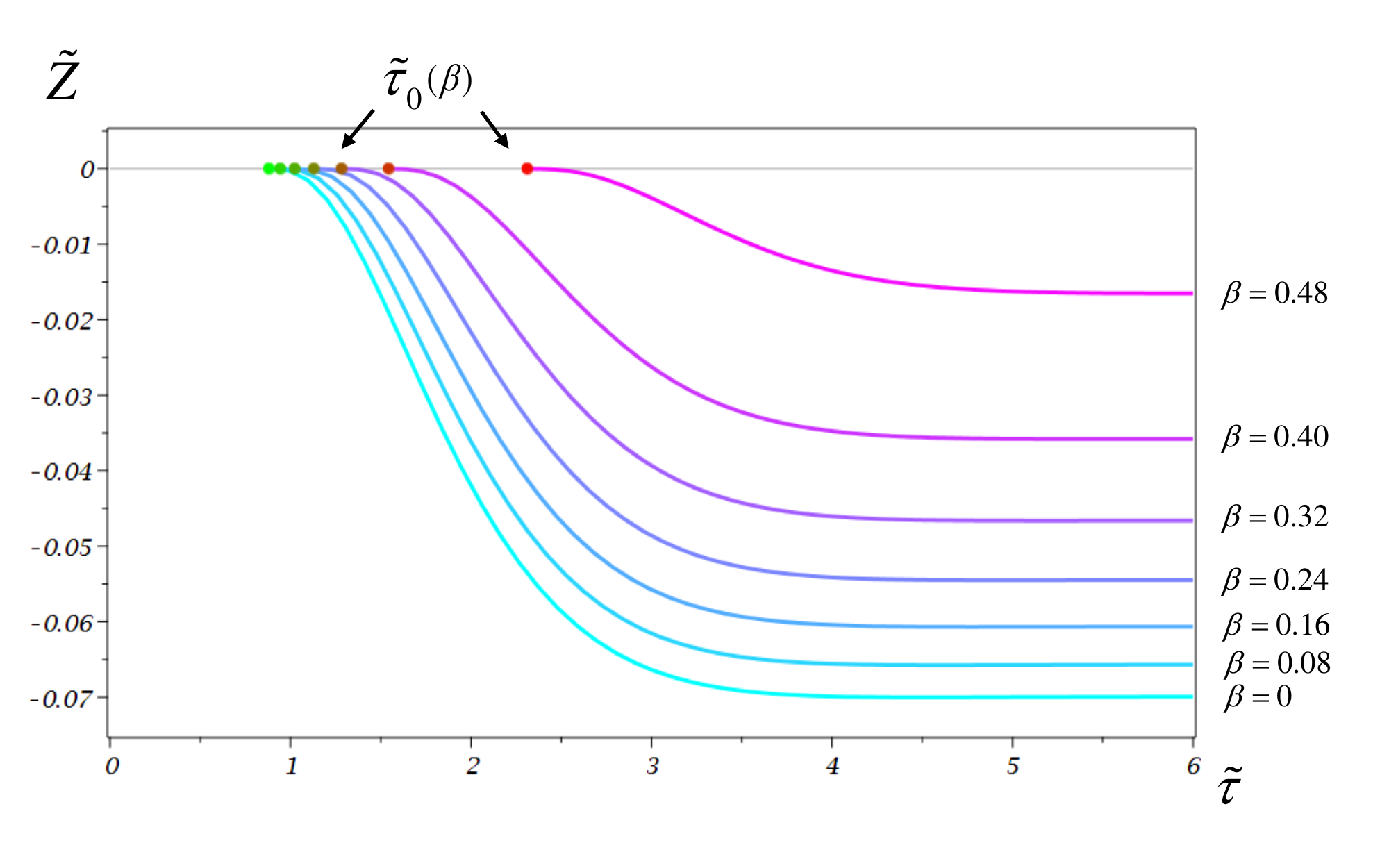}} &
\hspace{-0.15in}\imagetop{(b)} & \hspace{-0.1in}\imagetop{\includegraphics[width=0.45\textwidth]{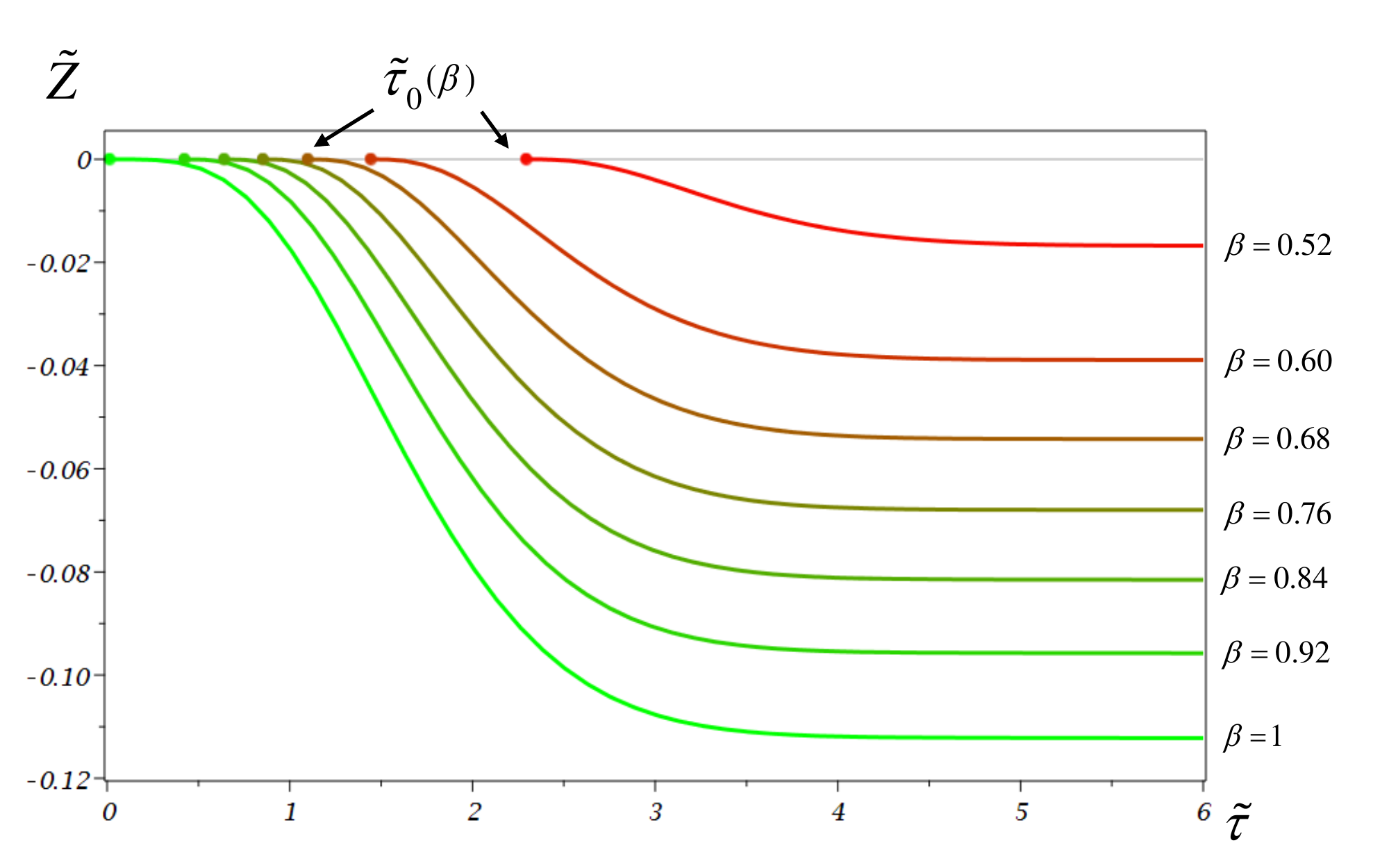}}
\end{tabular}
\caption{(a) $\tilde{Z}(\tilde{\tau})$ for the case $k=+1$, (b) $\tilde{Z}(\tilde{\tau})$ for the case $k=-1$, and various values of $\beta$.}
\label{Mpm}
\end{figure}

In order to demonstrate that a similar property is valid for the cases $k=\pm 1$ we numerically integrate the equation for $Z$. For this purpose we used the obtained earlier exact solutions for the dilaton field $\psi$  \eq{2A} and \eq{1B} and fixed the constant coefficients in the linear combination of the basic solutions by using  the initial conditions \eq{Zinit}.
Plots presented in figures \ref{Mpm}{\color{blue}a} and \ref{Mpm}{\color{blue}b} show $\tilde{Z}$ for different values of the dimensionless parameter $\beta$. For all the considered $\beta$ the function $\tilde{Z}$ (and hence $Z$) remains negative and does not pass through zero again. This supports our conclusion that once the de Sitter stage started it will last forever.
Let us note that for a fixed $\beta$ the function $\tilde{Z}$ at late time $\tilde{\tau}$ reaches a constant value. This ``saturation property" follows from the fact that the amplitude of the dilaton field exponentially decreases in this regime.


\section{Formation and evaporation of a  black hole}\label{SecVI}

\subsection{Vaidya-type generalization of the static solution}

In order to describe formation and evaporation of a  black hole in the limiting curvature theory of gravity we modify our model by including the matter into it.  For this purpose we add to action (\ref{LIM})  the following term
\be
I_m=-{1\over 2}\int d^2 |g|^{1/2} g^{\alpha\beta}\Phi_{,\alpha} \Phi_{,\beta} \, .
\ee
This action describes conformal massless scalar field $\Phi$. The corresponding field equation for this field is
\be \n{BP}
\Box \Phi=0\, .
\ee
This field gives a contribution
\be \n{DT}
\Delta {\cal T}_{\alpha\beta}=\Phi_{;\alpha} \Phi_{;\beta}-{1\over 2} g_{\alpha\beta}\Phi_{;\gamma} \Phi^{;\gamma}\,
\ee
to the "stress-energy" tensor $ {\cal T}_{\alpha\beta}$ which enters the gravity equations (\ref{eqchi1}).

Before we discuss how such a field $\Phi$ affects the vacuum solution which we described in the previous sections let us consider the equation (\ref{BP}) in a case when a background metric is of the form
\be \label{vr}
ds^2=-f dv^2+2dv\,dr\, ,
\ee
where $f$ is an arbitrary function of the coordinates $v$ and $r$. One has
\be \n{BV}
\Box \Phi\equiv 2{\partial^2 \Phi\over \partial v \partial r}+{\partial\over  \partial r}\left( f{\partial \Phi\over  \partial r}\right)=0\, .
\ee
This equation has a simple solution $\Phi=\Phi(v)$ which describes in-coming flux of radiation. Its stress-energy tensor is
\be \n{DTT}
\Delta {\cal T}_{\alpha\beta}=(\Phi_{,v})^2\ \delta_{\alpha}^v\delta_{\beta}^v\, .
\ee
This stress-energy tensor describes in-falling null fluid.
In the absence of the field $\Phi$ the metric of the subcritical solution in the advanced time coordinates $(v,r)$ has the form (\ref{vr}) with
\be \n{fff}
f=1-{m\over \lambda}\exp(-2\lambda r)\, .
\ee
One can check that if one adds a term (\ref{DTT}) to the gravity equations the only modification of the metric is that its parameter $m$ in (\ref{fff}) becomes a function $m(v)$ which obeys the equation (see e.g. \cite{Fabbri:2005mw} )
\be
{dm(v)\over dv}=(\Phi_{,v})^2\, .
\ee
The curvature of this $v-$dependent metric, which in fact is a Vaidya-type generalization of a static solution, is
\be
R=4m(v)\lambda \exp(-2\lambda r)\, .
\ee
The apparent horizon $H$ is determined by the condition
\be
\exp(-2\lambda r_\ins{H})={ \lambda\over m(v) }\, .
\ee
The equation for the junction surface inside the horizon where $R=\Lambda$ is
\be
\exp(-2\lambda r_\inds{\Lambda})={ \Lambda\over 4\lambda m(v) }\, .
\ee
For an arbitrary function $m(v)$ these two parameters obey the same equation \eq{rLrH} as in the case of a constant mass
\be
r_{-,\inds{\Lambda}}-r_\ins{H}={1\over 2\lambda}\ln \beta\, .
\ee

At the junction line $\Sigma$  one has
\be
f_\inds{\Lambda}=f(r_\inds{\Lambda})=1-\beta^{-1}\hh \beta={4\lambda^2\over \Lambda}\, .
\ee
As earlier we assume that $\beta <1$ so that the junction surface is located in the black hole interior.

\subsection{Simple model of an evaporating black hole}

In the above discussion we assumed that $\Phi$ is a classical field. For a quantum field $\Phi$ one can define the renormalized quantum average $\langle\Delta {\cal T}_{\alpha\beta}\rangle$ and insert it into the gravity equations. This object can be  reconstructed from the conformal anomaly. Such a model  was studied in details in \cite{Callan:1992rs}. A nice and comprehensive review of this model can be found in \cite{Fabbri:2005mw}.

We shall not discuss here quantum aspects of the problem. Instead of this we shall use known results concerning the Hawking radiation for such 2D black holes in order to choose a ``realistic" form of the mass function $m(v)$ in the above discussed classical Vaidya type solution. Let a black hole be formed by collapse of the null fluid and after some relaxation time it emits Hawking quanta.  As a result black-hole mass slowly decreases.

For a 2D conformal massless field one can obtain the quantum average of the stress-energy tensor by integrating the conformal anomaly \cite{Christensen:1977jc}. In appendix~\ref{AppC} we reproduce this derivation for the case of a static background metric and demonstrate that at the future null infinity the energy flux of the Hawking radiation generated by a black hole is exactly
compensated by the corresponding negative energy flux through the horizon.
In other words, the change of the  mass $m(u)$ registered by an external observer is accompanied by a similar mass change $m(v)$ at the black hole horizon.

 A detailed structure of the stress-energy of the quantum field near the horizon is quite complicated (see e.g. discussion in \cite{Abdolrahimi:2016emo,Bardeen:2020lko} and references therein).
In order to illustrate features of the Vaidya-type metric for an evaporating black hole we use the rate of change of its mass $m(v)$ which is consistent with the rate of its energy emission in the Hawking process
\footnote{
A simple model discussed in \cite{Hayward_2006,Frolov:2014jva}  illustrates how this relation between $m(u)$ and $m(v)$ arises. One can put a massive thin shell outside and close to the horizon of a quasistatic black hole which emits both outgoing positive energy flux to the infinity and negative energy flux through the horizon. The conservation law allows one to find parameters of such a shell. It was demonstrated that when the sum of these fluxes vanishes and the rate of the mass change is small, the stress-energy of the massive shell is also small and its backreaction on the metric can be neglected.
}.

Note that surface gravity $\kappa$ at the apparent horizon does not depend on the mass parameter and is given by
\be\label{kappa}
\kappa=\lambda .
\ee
It means that the Hawking temperature of the horizon for the evolving black hole described by the Vaidya metric \eq{vr} is constant during its evolution. If we would consider a quantum massless scalar field in the Unruh vacuum on this classical background, then the Hawking flux at infinity would be \cite{Callan:1992rs}
\ba\label{Tuu}
T^\ins{(Hawk)}_{uu}\big|_{\infty}=\frac{\hbar\lambda^2}{48\pi}.
\ea
For $N$ conformal fields the flux will be $N$ times larger. Thus the rate of evaporation of the black hole mass due to quantum effects is constant
\be
{dm(u)\over du}=-T^\ins{(Hawk)}_{uu}\big|_{\infty}\, .
\ee
This means that $m(u)$ is a linear function of the retarded time $u$. We can use the above discussed consistency condition and assume that $m(v)$ in the Vaidya type solution has a similar property, namely $m(v)$ is a linear function of the advanced time $v$.
This classical metric would qualitatively model the back reaction effect of quantum Hawking evaporation. This simplified model works well during the whole stage of evaporation except the last moment when $m(v)$ vanishes and the last impulse of energy might be emitted similar to the thunderbolt in RST model \cite{Russo:1992ax}.
In what follows we restrict ourselves to considering a simplified model in which the metric has Vaidya-type form and choose the mass-function $m(v)$ to mimic processes of a black hole formation and its further quantum evaporation.


\subsection{Null rays in the Vaidya type solution in the limiting curvature theory}

\subsubsection{In-coming and out-going null rays}

Discussion of properties of the Vaidya type version of 2D black holes is simplified if we make the following change of variables
\bea
&\ y=\exp(2\lambda r)\hh
\ \tilde{v}=\lambda v\, ,\\
& M(\tilde{v})=m(v)/\lambda\hh
dS^2=\lambda^2 ds^2\, .
\eea
Then one has
\bea
&dS^2=-f d\tilde{v}^2+{1\over y} d\tilde{v}\, dy \hh  f=1-{M(\tilde{v})\over y}\, .
\eea
The apparent horizon for this metric is defined by the condition $f(\V,y)=0$. Thus
\be
y_\ins{H}=M(\tilde{v})\, .
\ee
The equation for the junction line $\Sigma$ is
\be
y_{\Lambda}=\beta \, y_\ins{H}=\beta M(\tilde{v})\, .
\ee

If the rate of the mass change is given, one can use the above relations to find a position of the apparent horizon and junction line in $(v,y)$ coordinates. In the supercritical domain one can use the same coordinates $(\tau,t)$ as earlier. A form of the junction line as well as a relation between subcritical $(v,y)$ coordinates and supercritical ones follow from the matching conditions discussed in the appendix~\ref{AppA}. Namely, the  metric on the junction line $\Sigma$ induced by its embedding in both subcritical  and supercritical geometries should be the same and the extrinsic curvature is continuous. In a general case, the corresponding equations depend on $M(v)$ and are rather complicated. We shall not discuss them here.

\begin{figure}[!hbt]
\floatbox[{\capbeside\thisfloatsetup{capbesideposition={right,center},capbesidewidth=6cm}}]{figure}[\FBwidth]
{\caption{\small{Spacetime of 2D limited curvature black hole with the mass parameter $M(\tilde{v})$ which is approximately linear during the stage of the black hole formation and after a short period transitions to a slow linear evaporation stage. Before the BH formation and after its evaporation the null rays diverge exponentially $y\sim \exp{\tilde{v}}$.}}\label{RayTrace}}
{\includegraphics[width=0.45\textwidth]{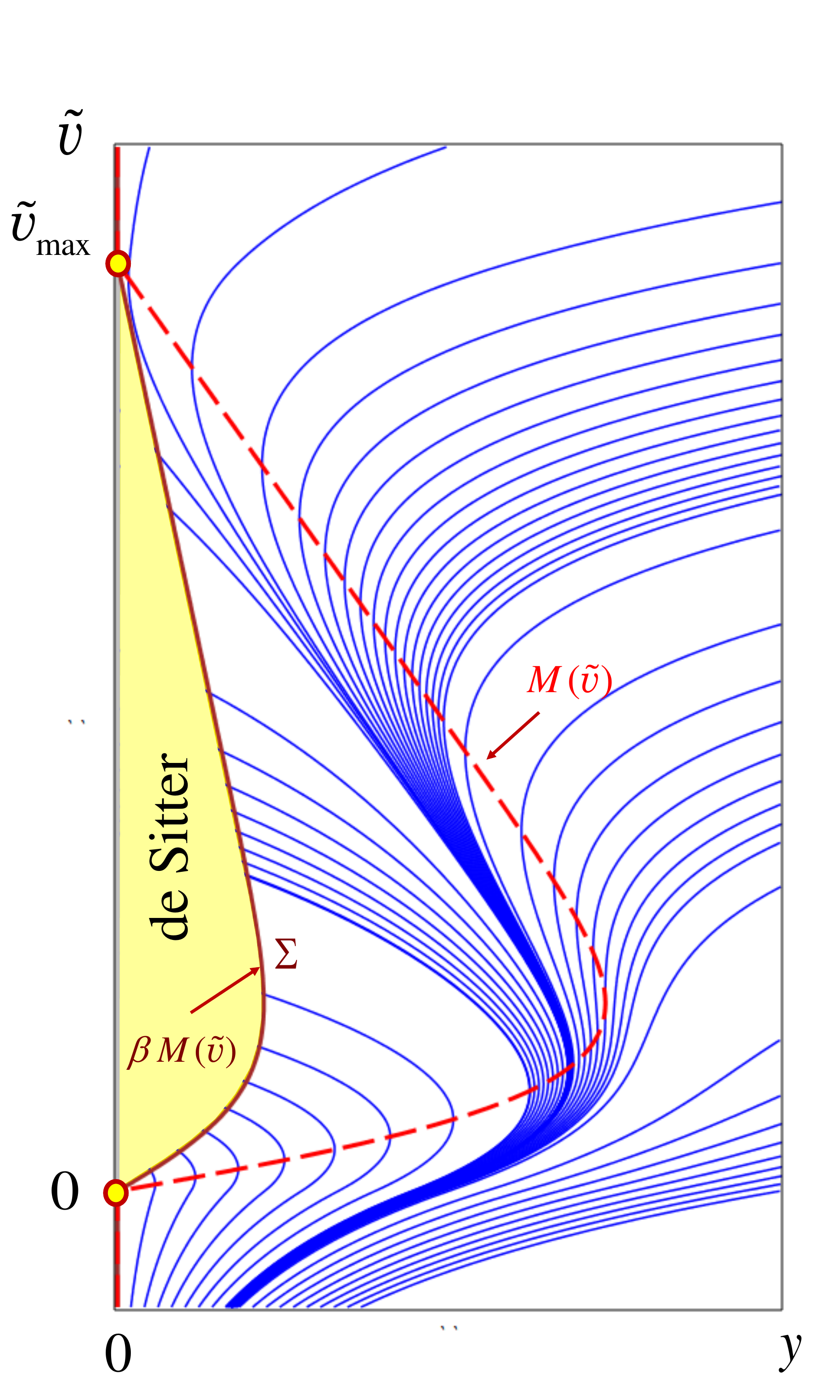}}
\end{figure}

In order to understand better causal structure of the Vaidya type solutions in the limiting curvature gravity it is instructive to consider null rays propagation. There exist two types of null rays:
\begin{itemize}
 \item Incoming rays that move along $v=$const line. Future directed null vectors  tangent to these rays are $k^{\alpha}=[0,-2y]$. These vectors are chosen so that $k_{\alpha}=-\V_{,\alpha}$;
 \item Outgoing rays. Their future directed tangent null vectors  are $l^{\alpha}=[2,2F]$.
 \end{itemize}
Here $F=y-M(\V)$. These vectors are normalized so that $(\ts{k},\ts{l})=-2$.

We denote by $\ts{q}_\ins{H}$ a vector
\be
{q}_\ins{H}{}_{\alpha}=F_{,\alpha}=[-M',1]\hh
M'={dM\over d\V}\, .
\ee
This vector is normal to $F=$const lines. Its sign is chosen so that the in-coming rays after crossing the apparent horizon enter the region with negative $F$
\be
(\ts{k},\ts{q}_\ins{H})|_{F=0}=-2y\, .
\ee
One also has
\be
(\ts{l},\ts{q}_\ins{H})|_{F=0}=-2M'\, .
\ee
This relation shows that at the phase when the mass $M$ increases the out-going null rays on the apparent horizon always enter the inner domain with $F<0$, while at the phase when the mass $M$ decreases, the out-going null rays can leave this domain.

These calculations can be easily repeated for the junction line obeying the equation $Q\equiv y-\beta M(\V)=0$. A vector normal to this line is $q^{\alpha}_\inds{\Lambda}=[-\beta M',1]$. One has
\be
(\ts{l},\ts{q}_\inds{\Lambda})|_{Q=0}=-2[\beta M'+(1-\beta)M].
\ee
This relation shows that when $M'>0$ the out-going rays always pass through the junction surface and enter the supercritical domain. For $M'<0$ some of these rays can leave the supercritical domain. This happens when the following condition is satisfied
\be \n{cond}
-{M'\over M}>{1\over \beta}-1\, .
\ee
At the phase when the rate of evaporation $-{d\ln M\over d\V}$ is small this does not occur. However at the final stage of the black hole evaporation when $M$ is small and $M'$ remains finite condition (\ref{cond}) might be satisfied. In such a case the "inner de Sitter core"  becomes visible to an external observer at the final stage of the black hole evaporation.

Figure~\ref{RayTrace} shows the qualitative behavior of the out-going null rays in the Vaidya-type model which describes creation and further evaporation of a black hole.

\subsubsection{Special case}

 We finish this section by the following useful remark. When the mass parameter in the Vaidya metric is either constant or it is a linear function of the advanced time $\V$ the outgoing null ray equations can be solved explicitely in terms of the elementary functions.

The equation for out-going ray when $M=M_0=\const$ is
\be
{dy\over d\V}=y- M_0\, .
\ee
Its solution is
\be
y=M_0+C \exp(\V)\, .
\ee

When the mass $M$ is a linear function of $\V$
\be
M=M_1+\alpha \V \, ,
\ee
one has
\be
{dy\over d\V}=y- M_1-\alpha \V\, .
\ee
By changing the variables
\be
y=M_1+\alpha \rho\,
\ee
one obtains the equation
\be
{d\rho\over d \V}=\rho-\V\, .
\ee
It can be easily integrated with the following result
\be
\rho=1+\V +C\exp(\V)\, .
\ee
It is interesting that in the 4D Vaidya metric when the mass is a linear function of the advanced time $v$ a solution of  equation for radial outgoing null rays can also be found analytically in terms of the elementary functions \cite{Volovich:1976aw}.

\section{Discussion}\label{SecVII}

The general relativity is incomplete theory both in the classical and quantum domains. In the classical Einstein gravity evidences of this incompleteness are cosmological and black-hole singularities predicted by the theory. To cure this problem one can look for a modification of the theory of gravity which forbids infinite growth of the curvature. In this paper we discuss such models. Our main idea is to incorporate the limiting curvature condition by choosing a special form of the covariant action. To construct it we use an approach which allows one to include the inequality constrains into the Lagrangian of a system by means of properly chosen Lagrange multipliers.

The dynamics  in such a modified theory  reduces to the motion of the system identical to the motion of the initial (unmodified) theory until the system reaches the regime where the inequality constraint(s) may be violated. After this it continues its motion along the constraint surface where the corresponding Lagrange multiplier becomes dynamical. To distinguish these two regimes we call the former one subcritical and the latter one supercritical. During the further evolution the system may remain on the constraint surface or return to its subcritical regime. If there are more than one constraints the system can also flip and continue its motion along another constraint surface.

Let us emphasize that this approach is novel and it has a number of appealing features:
\begin{itemize}
\item It is "geometric" in the sense that a theory is described by a covariant action which directly provides restrictions on the spacetime curvature. In the dilaton gravity models which have been used to obtain regular black hole solutions one needs additionally to check that the obtained solutions satisfy the limiting curvature condition;
\item We described application of the proposed approach for special case of two-dimensional black holes. However, the method itself allows a natural generalization to higher dimensional spacetimes.
\end{itemize}

In order to illustrate this approach we considered a well known model of 2D dilaton gravity \cite{Witten:1991yr,Callan:1985ia,Mandal:1991tz,Elitzur:1991cb,Frolov:1992xx,Dijkgraaf:1991ba,Callan:1992rs} and modified its action by adding a term which controls and limits the curvature growing. We solve the resulting equations of this covariant theory both in the subcritical and supercritical regime. We also derived the conditions which allows one to match these solutions at the junction surface where $R=\Lambda$. We constructed conformal Carter-Penrose diagram  for such an eternal black hole (see figure~\ref{CPD4}) and demonstrated that in our model instead of a singularity of the original black hole the solution of the limiting curvature model has two de Sitter cores. Inside the black hole in $T_-$ region it represents an expanding de Sitter universe.
In the spacetime of the eternal black hole a similar contracting de Sitter core is present in the white hole region $T_+$.
As it is expected the curvature of the obtained global complete spacetime is restricted and its value satisfies a condition $|R|\le \Lambda$. In the second part of the paper we analyzed two questions: (i) Can the obtained solution leave the constraint surface and return to its subcritical regime, and (ii) How to modify the obtained solution in order to describe black hole formation and its further evaporation.

To answer the first question we numerically solved the ordinary differential equations which allows one to trace the behavior of the corresponding control parameter $\chi$. We showed that once a solution reaches the supercritical regime it cannot return to the subcritical phase of its evolution.
To answer the second question we discussed a Vaidya-type generalization of the 2D black hole in the limiting curvature model. One of the obtained results  is that  at least in this model one can expect that the inner de Sitter core becomes visible to an external observer at the final stage of the black hole evaporation.

Certainly the 2D black hole model considered in the paper is oversimplified. It is chosen mainly because it
admits a quite complete analysis and exact analytical expressions for the metric and other fields in this model. However, the approach developed here is quite general and allows one to analyze more complicated black hole and cosmological models in four and higher dimensions.

Recently the limiting curvature gravity theory was applied for study of 4D cosmological models and it was shown that for a properly chosen curvature constraint such homogeneous isotropic universes have a bounce \cite{Frolov:2021}. Namely, after the curvature of a contracting universe reaches a critical value its solution becomes supercritical which universally possesses a turning point. For a "large" initial size of the universe its supercritical motion is very close to the evolution of the deSitter universe.

Another attractive feature is that the limiting curvature theory of gravity opens an interesting possibility to restrict growing of anisotropy by imposing restrictions of the Weyl tensor. This option is very important for study of the interior of four (and higher) dimensional black holes, as well in the bouncing cosmological  models. Certainly, the application of the limiting gravity models in four and higher dimensions requires additional work. There is a variety of curvature invariants which can be used to restrict the curvature of solutions. To find and to classify such constraints which properly restrict the curvature of solutions is an interesting problem.

In the models with the limiting curvature the constraint imposed on the curvature invariants is accompanied by Lagrange multiplier fields. In the two dimensional case which is considered in this paper these fields (as the metric) are not dynamical. In the higher dimensional case the control field $\chi$ (or its analogues) would obey nontrival equations. In this sense one can think that they introduce new degrees of freedom. It is well known that in the gravity models with higher in curvature terms a similar degrees of freedom corresponding to unphysical (ghost) modes are present. This results in a number of fundamental problems. It is an interesting problem to analyse the dynamical content of Lagrange multiplier fields in the four- and higher dimensional limiting curvature models. We mention here only one important generic  property of such fields. The field $\zeta$ is not dynamical in the sense that its value either vanish or simply is determined by the algebraic equation. The control field $\chi$ has zero value in the subcritical regime and the matching conditions uniquely determine its further evolution in the supercritical regime. In this sense it also does not have freely propagating modes.

These and other questions and applications the limiting curvature gravity approach to  cosmological and black hole problems require further study.  We hope to return to the discussion of such models and their properties in future publications

\appendix

\section{Matching conditions}\label{AppA}

In this appendix we discuss junction conditions for $\chi$ at the line $\Sigma$ separating subcritical and supercritical domains. For this purpose we use gravity equations (\ref{eqchi1}). Consider a two-dimensional spacetime with metric
\be
ds^2=g_{\alpha\beta}dx^{\alpha}dx^{\beta}\, .
\ee
Let $F(x^{\alpha})=0$ be an equation of $\Sigma$. We use the equation $F(x^{\alpha})=C$ to define a set of lines $\Sigma_C$ in the vicinity of $\Sigma=\Sigma_{C=0}$. We denote by $\ts{n}$  unit vectors orthogonal to $\Sigma_C$
\footnote{
The direction of the normal vector to the surface $\Sigma$ is chosen so that $n_{\alpha}=\epsilon\,\partial_\alpha F/|F^{;\sigma}F_{;\sigma}|^{1/2}$.
}
and by $\ts{e}$ unit vectors tangent to these lines. The following relations are valid
\be
(\ts{n},\ts{n})=-(\ts{e},\ts{e})=\epsilon\hhh
(\ts{n},\ts{e})=0\hhh e_{\alpha}=\varepsilon_{\alpha\beta}n^{\beta}\, .
\ee
Here $\epsilon=-1$ for spacelike lines $\Sigma_C$ and   $\epsilon=1$ for timelike lines. $\varepsilon_{\alpha\beta}$ is an antisymmetric tensor, $\varepsilon_{01}=\sqrt{|g|}$.

It is easy to check that
\ba
& n^{\beta}n_{\alpha ;\beta}=w e_{\alpha}\hhh
e^{\beta}e_{\alpha ;\beta}=H e_{\alpha} \hhh
 e^{\beta}n_{\alpha ;\beta}=H e_{\alpha}\hhh
 n^{\beta}e_{\alpha ;\beta}=w n_{\alpha}.
\ea
Here
\be
H=-\epsilon e^{\alpha}e^{\beta}n_{\alpha ;\beta}\hh
w=\epsilon n^{\alpha}n^{\beta}e_{\alpha ;\beta}\, .
\ee
The quantity $H$ has a meaning of the extrinsic curvature of the lines $\Sigma_C$, while (for the spacelike surfaces) $w$ is an acceleration of an observer with the velocity $\ts{n}$.
It is also easy to check that
\ba
&{\cal L}_{\ts{n}}\ts{e}=-{\cal L}_{\ts{e}}\ts{n}=[\ts{n},\ts{e}]\hh [\ts{n},\ts{e}]^{\alpha}=wn^{\alpha}-He^{\alpha}\, .
\ea
 Here ${\cal L}_{\ts{n}}\ts{e}$ is Lie derivative of vector $\ts{e}$ in the direction $\ts{n}$.

We denote
\be
D_n=n^{\alpha}\nabla_{\alpha}\hh D_e=e^{\alpha}\nabla_{\alpha}\, .
\ee
Then one has
\be\n{COM}
[D_e,D_n]\chi=-w D_n\chi +H D_e \chi\, .
\ee

To discuss junction conditions it is convenient to use scalar quantities. For this purpose we define for any  symmetric tensor ${\cal A}_{\alpha\beta}$ its components in the $(\ts{n},\ts{e})$ basis by means of the following relations
\ba
&{\cal A}_{nn}={\cal A}_{\alpha\beta}n^{\alpha} n^{\beta}\hh
{\cal A}_{ne}={\cal A}_{en}={\cal A}_{\alpha\beta}n^{\alpha} e^{\beta} \hh
{\cal A}_{ee}={\cal A}_{\alpha\beta}e^{\alpha} e^{\beta}\, .
\ea

For the tensor
\be
{\cal G}_{\alpha\beta}=\chi_{;\alpha\beta}-g_{\alpha\beta}\Box\chi-{1\over 2}g_{\alpha\beta}R\chi\, ,\n{j0}\\
\ee
one gets
\bea
&{\cal G}_{nn}=D_e^2\chi-H D_n\chi-{1\over 2}\epsilon R\chi\, ,\n{GNN}\\
& {\cal G}_{ee}=D_n^2\chi-w D_n\chi+{1\over 2}\epsilon R\chi\, ,\n{GEE}\\
&{\cal G}_{ne}=D_n D_e\chi-w D_n\chi =D_e D_n\chi-H D_e\chi\, .\n{GNE}
\eea
To obtain the last equality in the previous relation we used (\ref{COM}).

We denote by $B^{-}$ and $B^{+}$ a limit of a scalar function $B$ on $\Sigma$ taken in the subcritical and supercritical domains, respectively. We also denote by $[B]=B^{+}-B^{-}$ the jump of this function on  $\Sigma$. When $B$ is continuous its jump vanishes.
Using the above described matching conditions for $R$, $\psi$ and $\zeta$ one gets
\be \n{TT}
[{\cal T}_{nn}]=[{\cal T}_{ne}]=[{\cal T}_{ee}]=0\, ,
\ee
Hence the quantities ${\cal G}_{nn}$, ${\cal G}_{ee}$ and ${\cal G}_{ne}$ should be also continuous on $\Sigma$. We use this property to establish the required matching conditions.

Since $\zeta\ne 0$ in the subcritical domain equation (\ref{constr1}) gives
\be
\chi^-=(\psi^{-})^2\, .
\ee
We assume that $\chi$ is continuous on $\Sigma$.
The gravity equation which includes (\ref{GEE}) implied that $D_n\chi$ cannot have a jump on $\Sigma$. In the opposite case this equation would contain a $\delta-$dunction term which cannot be canceled. The gravity equation which includes (\ref{GNN}) guarantees that the extrinsic curvature $H$ is continuous on $\Sigma$.

\section{Conformal diagram  for the metric in the supercritical domain}\label{AppB}

\subsection{Conformal diagrams}

In this appendix we discuss Carter-Penrose conformal diagrams for the metric in the supercritical domain. We restrict ourselves by considering the case where $k=+1$ and $0<\beta<1/2$. The de Sitter metric in $(\tau,t)$ coordinates is
\ba
&ds_+^2=-d\tau^2+a^2 dt^2,  \hspace{2.1cm}    a=A_1 \cosh(\tilde{\tau}) ,\\
&A_1=\sqrt{{1\over 2\beta}-1},  \hspace{3cm} \tilde{\tau}=\sqrt{\Lambda\over 2}(\tau+\Delta\tau) ,\\
&\Delta\tilde{\tau} =\sqrt{\frac{2}{\beta}}\arccos\!\sqrt{\beta}-\arccoth(\sqrt{2(1-\beta)}).
\ea
Define  a new time coordinate $T$ such that  $dT=d\tau/a$. Integrating this relation one gets
\be
T=T_0+\frac{2}{\alpha} \arctan[\exp(\tilde{\tau})]\, .
\ee
Here $T_0$ is an integrating constant and
\be
{\alpha}=\sqrt{ {\Lambda\over 2}}A_1\, .
\ee
We use ambiguity in the choice of $T_0$ and write the expression for $a$ in the form
\be
a={A_1\over \cos({\alpha}T)}\, .
\ee
Combining the above results one obtains
\be \n{spp}
ds_+^2={A_1^2\over \cos^2({\alpha}T)}(-dT^2+dt^2)\, .
\ee

The conformal factor $\Omega_+$ infinitely grows when ${\alpha}T\to\pm \pi/2$. This form of the metric is very often used for construction of the Carter-Penrose conformal diagram for the 2D de Sitter spacetime with compact spatial dimension (see e.g. \cite{Spradlin:2001pw} and references therein).

\begin{figure}[!htb]
\centering \setlength{\labelsep}{0.0mm}
\begin{tabular}{l l l l}
\hspace{-0.05in}\imagetop{(a)} & \hspace{-0.065in}\imagetop{\includegraphics[width=2.75in]{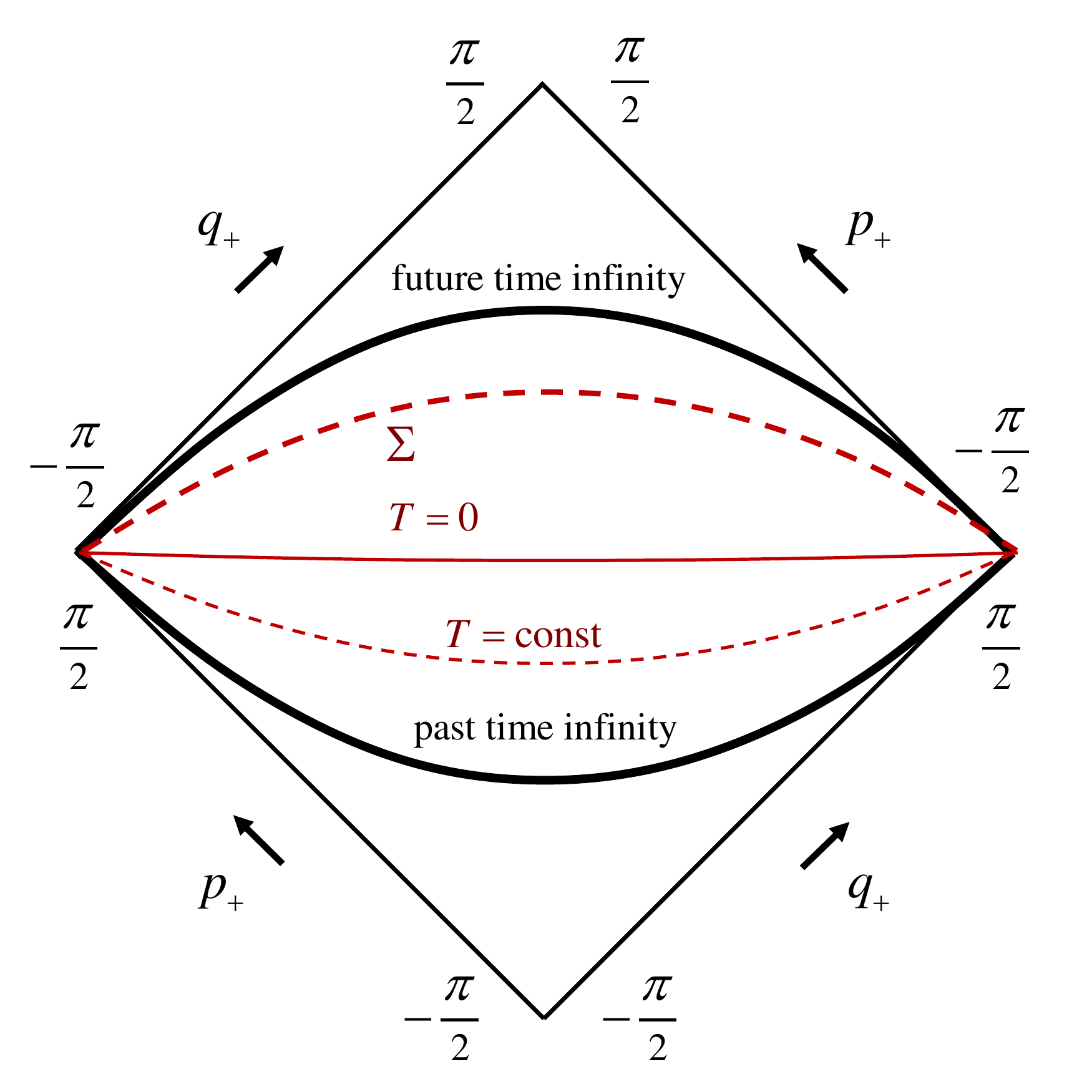}} &
\hspace{0.05in}\imagetop{(b)} & \hspace{-0.065in}\imagetop{\includegraphics[width=2.75in]{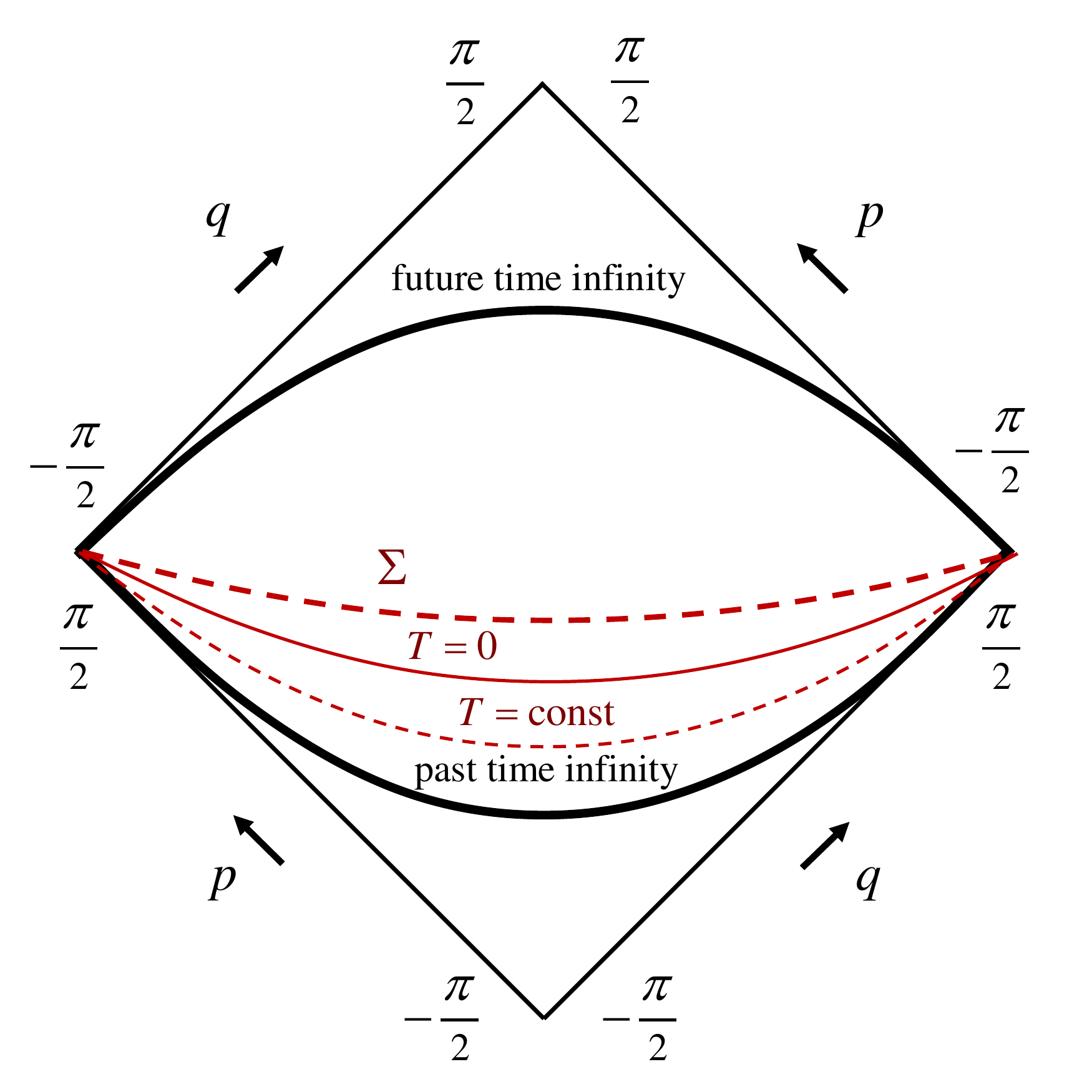}}
\end{tabular}
\caption{(a) Conformal diagram in $(p_+,q_+)$ coordinates. (b) Conformal diagram in $(p,q)$ coordinates}
\label{DS1}
\end{figure}

Since in our case $t\in (-\infty,\infty)$ for the construction of the conformal diagram an additional conformal transformation which brings the spatial infinity to a "finite coordinate distance" is required. This can be achieved by introducing new null coordinates $(p_+,q_+)$  that span the interval  $-\pi/2 < p_+, q_+ <\pi/2$,
\ba
&\tan(p_+)=T-t\hh
\tan(q_+)=T+t\, .
\ea
One obtains
\ba\n{SSPP}
& ds_+^2=-\Omega_+ dp_+\, dq_+ ,
\ea
where
\ba
&\Omega_+={A_1^2\over \cos^2 ({\alpha}T)\cos^2 p_+\cos^2 q_+}\hh
T={1\over 2}(\tan p_+ + \tan q_+)={\sin(p_+ + q_+)\over 2 \cos p_+ \cos q_+}.
\ea
Let us note that the above expressions are invariant under the discrete symmetry transformation $p_+\leftrightarrow q_+$.
The conformal diagram for the metric $ds_+^2$ in these coordinates is shown in figure~\ref{DS1}{\color{blue}a}. Two solid lines show curves where ${\alpha}T=\pm \pi/2$. They represent the future and past time infinity. Straight line $q_+=-p_+$ where $T=0$ separates two region with positive and negative values of $T$.
$T=$const curves are shown on  figure~\ref{DS1}{\color{blue}a}  by dashed lines.
 In the upper region where $T>0$ the extrinsic curvature parameter $H$ for a slice $T=\const$ is positive, while in the lower region it is negative.

The form of the metric (\ref{SSPP}) is preserved under the following change of the null coordinates
\be \n{PQ}
p=P(p_+)\hh q=Q(q_+)\, .
\ee
We assume that both functions $P$ and $Q$ are monotonic smooth functions of their arguments and these functions are chosen so that
\be
P(\pm\pi/2)=\pm\pi/2\hh Q(\pm\pi/2)=\pm\pi/2\, .
\ee
The conformal diagrams in the coordinates $(p,q)$ is qualitatively  the same as earlier. However, the shape and location of $T=$const line (including future and past infinities) would be shifted. Figure~\ref{DS1}{\color{blue}b} schematically shows the conformal diagram in $(p,q)$ coordinates.

\subsection{Matching subcritical and supercritical metrics}

In order to obtain a complete conformal diagram for 2D black hole in the limiting curvature model one needs to glue the diagram for the subcritical solution shown in figure~\ref{CPD} with two diagrams representing the de sitter cores. As it was mentioned earlier it is sufficient to do this for the core located in $T_-$ region.
For this purpose one needs to find functions $P$ and $Q$ which enter (\ref{PQ}). Let us describe how it can be done.

Let us consider first subcritical domain. Its metric in $(U,V)$ coordinates is (\ref{UV})
\be
ds^2=-{1\over \lambda^2}{dU\, dV\over 1-UV}\, .
\ee
The scalar curvature for this metric is
\be
R={4\lambda^2\over 1-UV}\, .
\ee
The junction surface $\Sigma$ in $(U,V)$ coordinates has the form
\be
UV|_{\Sigma}=1-\beta\, .
\ee
One can write this equation in the following parametric form
\be
U=\sqrt{1-\beta}\,\sigma\hh V=\sqrt{1-\beta}\,{1\over \sigma}\, .
\ee
The induced metric on $\Sigma$ is
\be \n{inm}
dl^2={1-\beta\over \beta} {d\sigma^2\over \lambda^2 \sigma^2}\, .
\ee
We write the metric in the supercritical domain in the form
\be
ds_+^2=-{A_1^2\, du_+\, dv_+\over \cos^2(bT)}\, ,
\ee
where
\ba
&u_+=T-t \hh
v_+=T+t \hh
A_1=\sqrt{{1\over 2\beta}-1}\hh
b=\sqrt{\Lambda\over 2}\, A_1\, .
\ea
The junction surface $\Sigma$ in $(u_+,v_+)$ coordinates has the form
\be
(u_+ +v_+)|_{\Sigma}=2T_{0}\hh   \cos(bT_0)=\sqrt{{1-2\beta\over 2(1-\beta)}}\, .
\ee
The parameter $T_0$ in this relation is singled out by the property that the extrinsic curvature is continuous on $\Sigma$.
This equation for $\Sigma$ can be written in the following parametric form
\be \n{inp}
u_+=T_0+\sigma_+\hh v_+=T_0-\sigma_+\, .
\ee
The induced metric on $\Sigma$ is
\be\n{inp}
dl_+^2={1-\beta\over  \beta} d\sigma_+^2\, .
\ee
The condition that the induced metrics calculated on $\Sigma$ from both sub- and supercritical sides are identical gives the following relation between $\sigma$ and $\sigma_+$
\be
\sigma_+={1\over \lambda}\ln \sigma\, .
\ee
Thus one has
\bea
&u_+=\hat{T}_0-{1\over \lambda} \ln U  \hh
v_+=\hat{T}_0+{1\over \lambda} \ln V   \hh
\hat{T}_0=T_0 -{1\over 2\lambda}\ln(1-\beta)\, .
\eea
Using these relations one can find the functions $P$ and $Q$ which enter the relation (\ref{PQ}).

\section{Conformal anomaly and the energy-momentum fluxes}\label{AppC}

Let us consider a static 2D metric
\be
ds^2=-f dt^2+{dr^2\over f}\hh f=f(r)\, .
\ee
We assume that $f$ vanishes at some point $r=r_H$ and $df/dr|_{r_\ins{H}}$ is finite.
Surface gravity at the horizon is given by
\ba
\kappa=\frac{1}{2}\frac{df}{dr}\Big|_{r_\ins{H}}.
\ea
We also assume that $f$  is positive and monotonically decreasing for $r>r_\ins{H}$ and $f(r=\infty)=-1$.

We denote
\vspace{-0.5cm}
\bea
&v=t+r_*\hh u=t-r_*\hh dr_*={dr\over f}\, ,\\
&k_{\alpha}=-v_{,\alpha}=(-1,-1/f)\hh
l_{\alpha}=-f u_{,\alpha}=(-f,1)\, .
\eea
Vectors $\ts{k}$ and $\ts{l}$ are null and satisfy the condition $(\ts{l},\ts{k})=-2$. These vectors are chosen so that they are regular at the future horizon.

We define
\be
T_{\alpha\beta}=\rho_+l_{\alpha}l_{\beta}+\rho_-k_{\alpha}k_{\beta}+{b\over 2}R\, g_{\alpha\beta}\, .
\ee
Here $R$ is the curvature of the metric
\be
R=-{d^2f\over dr^2}\, .
\ee
One has
\be
T_{\alpha}^{\alpha}=bR\, .
\ee
We identify $T_{\alpha\beta}$ with the quantum average of the stress-energy tensor of a conformal massless field.
In this case the coefficient $b$ is defined by the conformal anomaly.
For a single scalar field the coefficient $b=\hbar/(24\pi)$.
We assume that $T_{\alpha\beta}$ is stationary, that is ${\cal L}_{\xi}\ts{T}=0$, where  ${\cal L}_{\xi}$ is a Lie dervative along the Killing vector $\ts{\xi}=\partial_t$. This condition implies that $\rho_{\pm}$ depend only of $r$.
The conservation law $T^{\alpha\beta}\!{}_{;\beta}=0$ gives the following expressions for $\rho_{\pm}$
\ba
& \rho_+={1\over f^2}(S+C_+)  \hh
\rho_-=S+C_-   \hh
S=-{b\over 8}\big[ \big(\frac{df}{dr}\big)^2-2 f \frac{d^2f}{dr^2}\big]\, .
\ea
A condition that there is no incoming flux from the past null infinity fixes the integration constant $C_-=0$. Hence at the horizon
\be
\rho_{-,\ins{H}}=-{b\over 8} \Big( {df\over dr}\Big)^2=-{b\over 2}\kappa^2\, .
\ee
Regularity of $\rho_+$ at the horizon gives
\be
C_+={b\over 8} \Big( {df\over dr}\Big)^2={b\over 2}\kappa^2\, .
\ee
This shows that the positive energy flux of the Hawking radiation at the infinity ${b\over 2}\kappa^2$ (see e.g. Eq.\eq{Tuu}) is accompanied by the negative energy flux  $\rho_{-,\ins{H}}=-{b\over 2}\kappa^2$ through the horizon.


\section*{Acknowledgments}

The authors thank the Natural Sciences and Engineering Research Council of Canada and the Killam Trust for their financial support.





\providecommand{\href}[2]{#2}\begingroup\raggedright\endgroup

\end{document}